\documentclass[11pt, a4paper]{article}
\pdfoutput=1
\usepackage[T1]{fontenc}
\usepackage[utf8]{inputenc}
\usepackage[english]{babel}
\usepackage{amsmath}
\usepackage{amssymb}
\usepackage{graphicx}
\usepackage{xcolor, color}
\usepackage{url}
\usepackage{xspace}
\usepackage{subcaption}
\usepackage{jcappub}
\usepackage{natbib}

\definecolor{light-gray}{gray}{0.70}

\title{Red, Straight, no bends: primordial power spectrum reconstruction  from CMB and large-scale structure}
\author[a,b,c]{Andrea Ravenni}
\author[c,d,e,f]{Licia Verde}%
\author[c]{and Antonio  J.  Cuesta}
\affiliation[a]{Dipartimento di Fisica e Astronomia ``G. Galilei'', Universit{\`a} degli Studi di Padova, via Marzolo 8, I-35131, Padova, Italy}
\affiliation[b]{INFN, Sezione di Padova, via Marzolo 8, I-35131, Padova, Italy}
\affiliation[c]{Institut de Ci{\`e}ncies del Cosmos (ICCUB), Universitat de Barcelona (IEEC-UB), Mart{\'\i} i Franqu{\`e}s 1, E08028 Barcelona, Spain}
\affiliation[d]{ICREA (Instituci\'o catalana de recerca i estudis avan\c{c}ats)}
\affiliation[e]{Radcliffe Institute for Advanced Study, Harvard University, MA 02138, USA}
\affiliation[f]{Institute of Theoretical Astrophysics, University of Oslo, 0315 Oslo, Norway}
\emailAdd{andrea.ravenni@pd.infn.it}
\emailAdd{liciaverde@icc.ub.edu}
\emailAdd{ajcuesta@icc.ub.edu}
\abstract{We present a minimally parametric, model independent reconstruction of the shape of the primordial power spectrum. Our smoothing spline technique is well-suited to search for smooth features such as deviations from scale invariance, and deviations from a power law such as running of the spectral index or small-scale power suppression. We use a comprehensive set of the state-of the art cosmological data: {\it Planck} observations of the temperature and polarisation  anisotropies of the cosmic microwave background, WiggleZ and Sloan Digital Sky Survey Data Release 7 galaxy power spectra and  the Canada-France-Hawaii Lensing Survey correlation function. 
This reconstruction strongly supports the evidence for a power law primordial power spectrum with a red tilt and disfavours  deviations from a power law power spectrum including small-scale power suppression such as that induced by significantly massive neutrinos. This offers a powerful confirmation of the inflationary paradigm, justifying the adoption  of the  inflationary prior in cosmological analyses.
}
\begin{document}
\maketitle
\flushbottom

\section{Introduction}
 \label{sec:intro}
All recent cosmological observations are in excellent agreement with the standard $\Lambda$CDM model: a spatially flat cosmological model, with matter-energy density dominated by a cosmological constant  and cold dark matter, where cosmological neutrinos are effectively massless and where the primordial power spectrum of adiabatic perturbations is a (almost scale invariant)  power law.
State-of-the art cosmological observations such as those of  the \textit{Planck} satellite \cite{Planck:Overview}, measuring cosmic microwave background (CMB) anisotropies, provided us with very precise  measurements of the parameters of the standard cosmological model \cite{Planck:CosmologicalParameters2015}.

Most cosmological analyses assume a power-law primordial power spectrum  with a fixed spectral index, and deviations from this assumption are often in the form of a ``running'' of the spectral index.

A nearly scale invariant power spectrum is a generic prediction of the simplest models of inflation, but there are models with (small) deviations from this prediction (e.g., \cite{Miranda:Steps, Meerburg:Oscillations, Chen:StandardClock, Danielsson:Transplanckian}). Small deviations from scale invariance constitute  a critical  and generic prediction of inflation.  For this reason a model-independent reconstruction of the primordial power spectrum (PPS) shape  can be a powerful test of inflationary models.

Here we  perform a minimally parametric reconstruction of the PPS using smoothing spline interpolation in combination with cross validation. This approach follows \cite{Sealfon:reconstruction, Verde:MinimallyParamRec2008, Peiris:bPlanck2010}.

The idea is simple: we choose a functional form that allows a great deal of freedom in the shape of the deviations from a power-law.
Because most models predict the PPS to be smooth,  among the possible choices we use a smoothing spline.
The ensuing challenge is to avoid over-fitting the data;  a complex function that fits the data set extremely well is of no interest if  we are simply fitting statistical noise.  
To prevent over-fitting we use cross-validation and a roughness penalty.
The  roughness penalty is an additional  parameter that penalises a high degree of structure in the functional form.
By performing cross-validation as a function of this penalty, we can judge the amount of freedom in the smoothing spline that the data require, without fitting the noise.

The \textit{Planck} collaboration has performed an analysis with the same goals in mind, but with different methods \cite{Planck:ConstraintsOnInflation}.
They carried out both a parametric search for deviations from a power law, using a set of theoretically motivated shapes for the PPS, and a minimally parametric analysis to reconstruct the PPS.
In all cases there is no strong evidence for deviations from a power law.

Our analysis differs from that of the  \textit{Planck} collaboration and from others existing in the literature  as we analyse jointly a comprehensive set of state-of-the-art experiments probing the matter power spectrum and the latest \textit{Planck}  measurements.

Because we assume standard late-time evolution of density perturbations and consider both early-time observables (CMB) and late-time ones (i.e., large-scale structure), our reconstruction is  also  sensitive to late-time effects on structure formation. In particular a non-negligible neutrino mass would suppress the growth of structures below the neutrino free-streaming scale, inducing an ``effective'' loss of small scale power in our reconstructed PPS.
Reconstructing  in a model-independent way a possible neutrino signature on the shape of the matter power spectrum is of particular importance as 
\cite{Verde:CMBLocalH0, BattyeMoss,WymanRudd,HamannHasenkamp,BeutlerSaitoCuesta,GiusarmaDiValentino} claims that relatively large  neutrino masses ($\Sigma_{\nu} \gtrsim 0.4$ eV) could solve the tension between CMB and local measurements, whilst other studies \cite{Planck:CosmologicalParameters,Verde:NeutrinoEvidence?,HuCai,Verde:DarkRadiationEvidence,Efstathiou:H0revisited,Verde:noconcordance, Cuestanu, PalanqueDelabrouille}  rule out this possibility.

The rest of the paper is organised as follows: in section~\ref{sec:methods} we briefly summarise the methodology adopted, the data chosen and how they are analysed. In section~\ref{sec:results} we present our findings; we discuss and present the conclusions in section~\ref{sec:conclusions}.
\section{Methodology and datasets}
\label{sec:methods}
\subsection{Spline reconstruction}
We perform a minimally-parametric reconstruction of the primordial power spectrum based on the method  presented in  \cite{Sealfon:reconstruction} and further refined in \cite{Verde:MinimallyParamRec2008, Peiris:bPlanck2010}. Here we only briefly summarise the approach; it is based on the  cubic smoothing spline technique (for details see  \cite{Green:Nonparametric}).  In this approach 
to recover a smooth function $f(x)$, given its value $f_i$ only on a set of $n$ points $x_i$, hereafter \textit{knots}, one fits the pairs $(x_i, f_i)$ with  a cubic spline $s(x)$.
The spline, its first, and second derivatives are continuous on the knots by definition.
The full function is then uniquely defined by the values at  the knots and two boundary conditions.
We choose to require that the jump in the third derivative across the first and last knots is forced to zero.

In our application the resulting spline function  is the reconstructed primordial power spectrum. The $f_i$  are free parameters we wish to determine and  we
place the knots equally spaced in $\log k$ as it is the most conservative choice  to recover deviations from a power law.
The whole $s(k)$ is used as the PPS to compute the observables and evaluate the likelihood of the parameters $f_i$.
Including the roughness penalty, the  effective likelihood becomes 
\begin{equation}
-\log(\mathcal{L})=-\log(\mathcal{L}_{\rm exp}) + \alpha_p \int^{\ln{k_f}}_{\ln{k_i}} \left( s''(\ln{k})\right)^2 \, d \ln{k}
\label{penalty}
\end{equation}
where  $s''$ denotes the second derivative of $s$ with respect to $\ln k$, $k_i$ and $k_f$ are respectively the position of the first and of the last knots, $\alpha_p$ is a weight that controls the penalty, and $\mathcal{L}_{\rm exp}$ is the likelihood given by the experiments.

The roughness penalty effectively reduces the degrees of freedom, disfavouring jagged functions that ``fit the noise''.
As $\alpha_p$ goes to infinity, one effectively implements linear regression; as $\alpha_p$ goes to zero one is interpolating.
The use of cubic spline --- instead of other possible interpolating functions --- is motivated by the fact that such a function minimises the roughness penalty for a given set of knots $(f_i,x_i)$.

In generic applications of smoothing splines, cross-validation is a rigorous statistical technique for choosing the optimal roughness penalty \cite{Green:Nonparametric}.
Cross-validation (CV) quantifies the notion that if the PPS has been correctly recovered, we should be able to accurately predict new independent data.
To make the problem computationally manageable,
we adopt the following. We split the data set in two halves $A$ and $B$.
A Markov chain Monte Carlo (MCMC) parameter estimation analysis (for a given roughness penalty) is carried out on $A$, finding the best fit model.
Then the $- \log$ likelihood of $B$ given the best fit model for $A$, $CV_{AB}$, is computed and stored.
This is repeated by switching the roles of the two halves, obtaining $CV_{BA}$.
The sum $CV_{AB}+CV_{BA}$, gives the ``CV score'' for that penalty weight.
With this construction, the smoothing parameter that best describes the entire data set is the one that minimises the CV score.
The cross validation data sets are described below (see table \ref{tab:CV}).

We choose to use 5 knots equally spaced in $\log k$ between $k = 10^{-5}$ Mpc$^{-1}$ and $k=1$ Mpc$^{-1}$, i.e., $k_1=10^{-5}$ Mpc$^{-1}$, $k_2=1.78\times 10^{-4}$ Mpc$^{-1}$, $k_3=3.16\times 10^{-3}$ Mpc$^{-1}$, $k_4=5.62\times 10^{-2}$ Mpc$^{-1}$, $k_5=1$ Mpc$^{-1}$ (see figure 1 bottom panel for knots placement visualisation).
The number and position of the knots is  held fixed throughout the analysis. 
As discussed in reference \cite{Verde:MinimallyParamRec2008}, beyond a minimum number of knots, there is a trade-off between the number of knots and the penalty, and the form of the reconstructed function does not depend significantly on the number of knots beyond this minimum number.
As the main goal of this work is to explore, in a minimally parametric way, smooth deviations from a power law, a few ($> 3$) knots are sufficient.

The basic cosmological parameters, $\omega_b=\Omega_b h^2$, $\omega_c=\Omega_ch^2$, $h$, and $\tau_{\rm{reio}}$ --- physical baryonic matter density parameter, physical cold dark matter parameter, dimensionless Hubble parameter and optical depth to last scattering surface ---  are varied in the MCMC alongside the values $f_i$ of  the reconstruction at the knots. A flat geometry is assumed so that $\Omega_m+\Omega_{\Lambda}=1$.

The  prediction for cosmological observables, the calculation of the likelihood and the MCMC  parameter inference are  implemented using the standard Boltzmann code CLASS \cite{Lesgourgues:CLASS} and its Monte Carlo code, Monte Python (MP) \cite{Audren:MP}, suitably modified.\footnote{\url{http://class-code.net}}\footnote{\url{http://baudren.github.io/montepython.html}}

Even though we reconstruct the primordial power spectrum, we are sensitive to late-time cosmological effects. Our main focus is on massive neutrinos: the presence of non-negligibly  massive neutrinos would distort our reconstruction in a way that is predictable due to the linearity of the growth functions \cite{Dodelson:Cosmology} (see Appendix). Thus in the analysis we will assume massless neutrinos.

Of course neutrino masses do not actually affect the physical PPS. But assuming standard gravity, standard growth of structure, and massless neutrinos in the analysis, would yield a reconstructed PPS with an artificial distortion, if neutrino masses were not negligible. In fact a detectable signature of massive neutrinos in the real data would  appear as a  power suppression feature in the reconstructed PPS.
Of course a detection of power suppression cannot be univocally interpreted as signature of neutrino masses; other particles beyond the standard model could easily share the same properties of neutrinos when it comes to damping perturbations or it could be a real feature in the PPS.

\subsection{Datasets}
\label{sec:datasets}
\begin{figure}
\centering
\includegraphics[width=\textwidth]{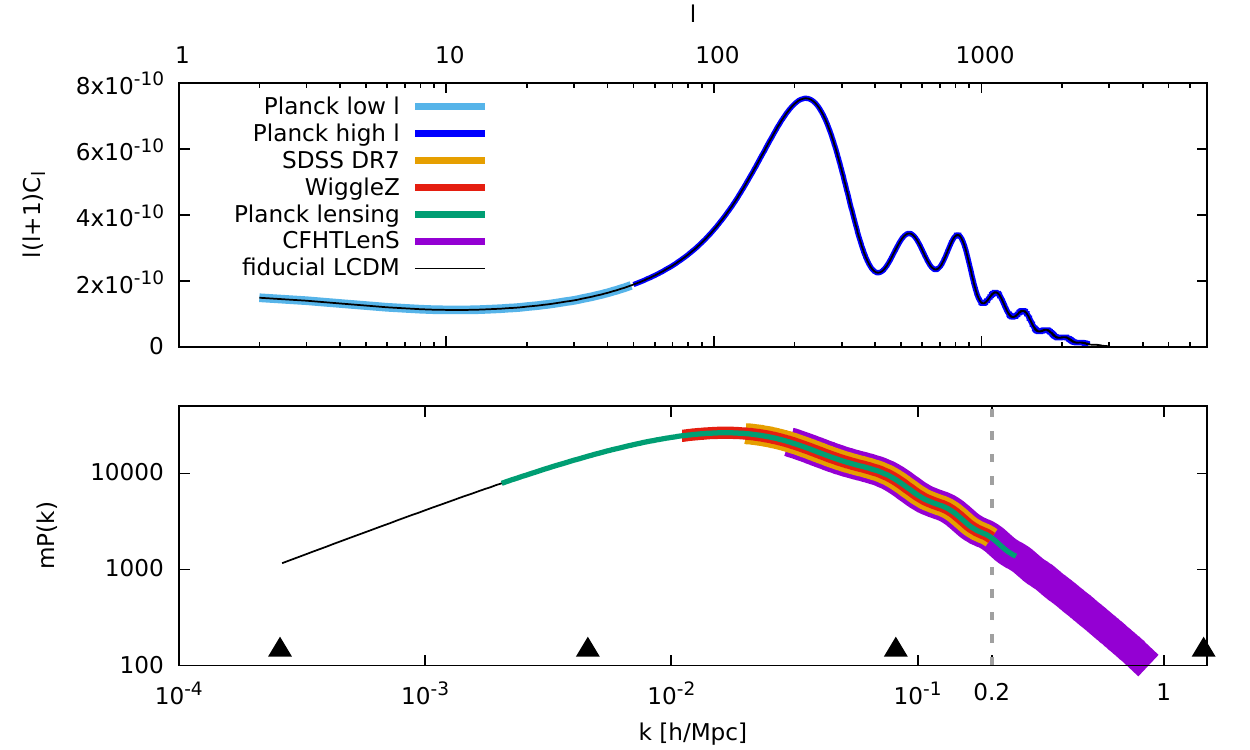}
\caption{Comoving scales covered by the experiments used in our analysis. The vertical dashed line show the limit of the linear scales. The triangles show the position of the knots. The leftmost one is not visible in the plot.}
\label{fig:CoveredScales}
\end{figure}

We use a comprehensive set of power spectra obtained from observations of CMB and of large scale structure (including both weak gravitational lensing and galaxies redshift surveys) as follows:
\begin{itemize}
\item \textit{Planck} power spectra of temperature and polarisation of the CMB. 
The \textit{Planck} collaboration released in 2013 the temperature data from the first half of the mission  \cite{Planck:CMBPowerSpectra}.
We complement the {\it Planck} 2013 data with the  WMAP polarisation. We refer to this as  PlanckCMB13. 
In 2015 the results of the full analysis has been released \cite{Planck:CMBPS15}.
Temperature and E-mode polarisation power spectra (and their cross-correlation) data and likelihoods come in two sets: a {\it low $\ell$}   from  $\ell=2$ to $\ell=49$, and the {\it high $\ell$} angular power spectrum.
We use the temperature and polarisation data up to $\ell =2500$ and we refer to this as  PlanckCMB15. 

\item Beside the CMB power spectrum, \textit{Planck} reconstructed the CMB lensing potential \cite{Planck:lensing}, which contains information on the amplitude of large scale structure integrated from recombination to present time. We will refer to it as PlanckLens.

\item The Canada-France-Hawaii Lensing Survey  (CFHTLenS) \cite{Heymans:CFHTLenS} provides the two point correlation function of the tomographic weak lensing signal.

\item  The WiggleZ Dark Energy Survey (WiggleZ), through the measurement of  position and redshift  of 238,000 galaxies,  mapped a volume of one cubic gigaparsec over seven regions of the sky up to a redshift $z \lesssim 1$.
The  corresponding galaxy power spectrum is presented in \cite{Parkinson:WiggleZFInalData}.

\item  The Sloan Digital Sky Survey collaboration, in Data release 7 (SDSS DR7), used a sample of luminous red galaxies to reconstruct the halo density field and its power spectrum roughly between $k=0.02$ $h$/Mpc and $k=0.2$ $h$/Mpc \cite{SDSS:DR7}.
\end{itemize}

In figure \ref{fig:CoveredScales} we show the scales probed by each experiment along with the location of the knots.

\subsection{Runs set-up}
\begin{figure}
\centering
\includegraphics[width=0.7\textwidth]{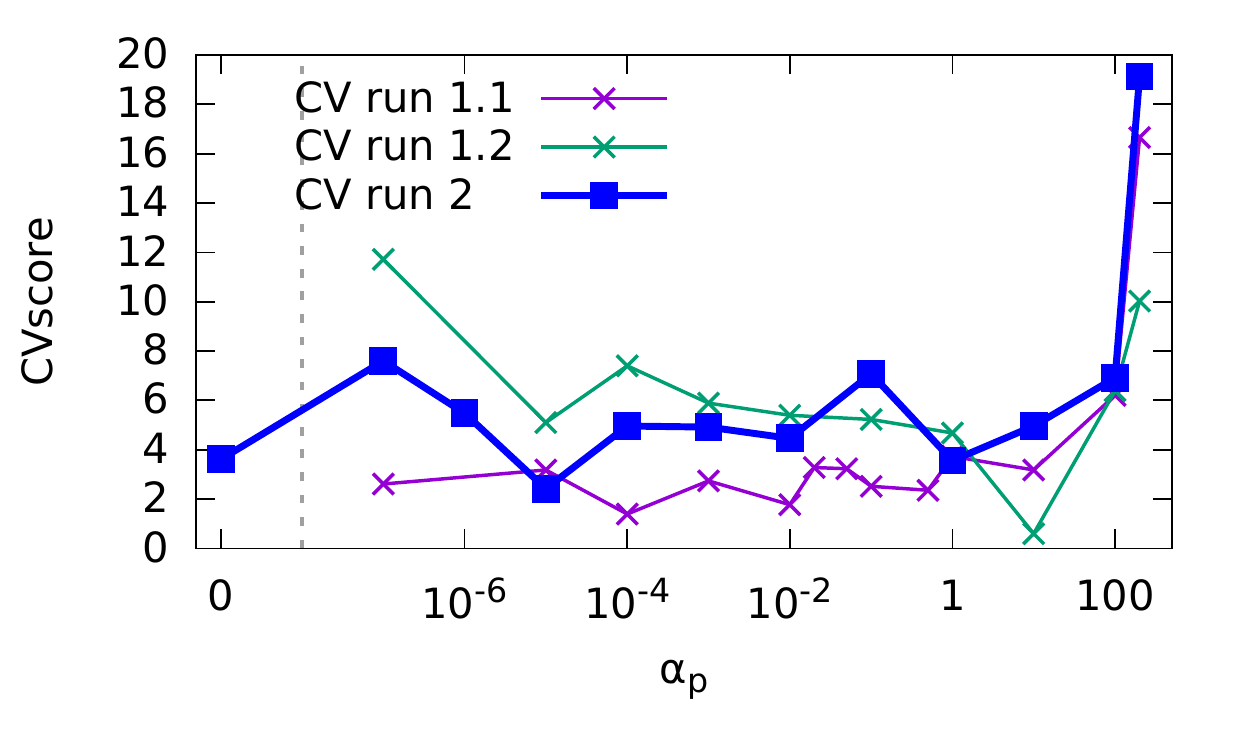}
\caption{CV score as a function of $\alpha_p$ for the cross-validation run 2. A different arbitrary offset has been subtracted from each CV score.}
\label{fig:CVscore2}
\end{figure}

We now describe the cross validation set up. 
In order to constrain both the shape of the PPS and the cosmological parameters,  we have to consider CMB primary data in all CV runs. Because of time constraints PlanckCMB2013 is used in the set up CV runs but PlanckCMB2015 is used in the final run. This choice is conservative, favouring slightly  more freedom (lower penalty) to the reconstructed PPS.
 Besides these, we have 4 other experiments: 2 measuring weak lensing and 2 using galaxy catalogues. We  perform  3 CV runs in a pyramidal scheme as summarised in table \ref{tab:CV}. We start performing in parallel two different cross-validation analysis on two pairs of experiments where each pair is formed by a weak lensing experiment and by a galaxy catalogue. The dependence of the  CV score on $\alpha_p$ was  mapped by sampling several  $\alpha_p$ values.
The results of these preliminary runs show no unexpected behaviour or tension, i.e., the reconstructed PPS shows no significant  deviation from a power-law, and the shape of the CV score is the same for both run 1.1 and run 1.2.
Knowing this, we then combine the large scale structure data to have one weak lensing and one galaxy survey in each CV set. The best roughness penalty  found from this CV is used in the final run which includes all experiments (this is called ``Rec.'' run in the table).
The  penalty parameter value to use in the reconstruction is determined by the CV score of run 2 alone: its dependence on $\alpha_p$  is   illustrated in figure~\ref{fig:CVscore2}.
The fact that the shape of the three CV scores --- from run 1.1, 1.2, and 2 --- shown in figure \ref{fig:CVscore2} is very similar,  indicate robustness and  that there are no  significant tensions between the  datasets.

\begin{table}[tbp]
\begin{tabular}{|ccc|}
\hline
Run		&$A$		&$B$	\\
\hline
1.1		&PlanckCMB13, PlanckLens				&PlanckCMB13, SDSS DR7	\\
1.2		&PlanckCMB13,	CFHTLenS				&PlanckCMB13, WiggleZ	\\
\hline
2		&PlanckCMB13,	PlanckLens, SDSS DR7	&PlanckCMB13, CFHTLenS, WiggleZ\\
\hline
Rec.    & \multicolumn{2}{c|}{PlanckCMB15, PlanckLens, SDSS DR7, CFHTLenS, WiggleZ}\\
\hline
\end{tabular}
\caption{Cross-validation datasets A and B for the various runs. The reconstruction (Rec.) involve all the experiments together.}
\label{tab:CV}
\end{table}

The CV score has a fairly well defined ``wall'' for  high penalties , but is quite constant under a certain threshold at $\alpha_p \sim 10$.
For high $\alpha_p$ the penalty  starts being the dominant  contribution to the likelihood, so the behaviour  in the limit of  high  $\alpha_p$ is expected. On the other hand, if small values of the penalty were to lead to overfitting, the CV score should increase as $\alpha_p$ decreases.  This is not what we see and can be understood as follows. CMB angular power spectra are always included in the analysis and  in this limit, it is the statistical power of these data (not the penalty) that drives  the smoothness of the reconstruction and therefore the  CV score.
In other words, for low values of the penalty below $\alpha_p\sim 10$, all datasets are  well consistent with the Planck-inferred PPS reconstruction: the CMB data alone disfavour unnecessarily wiggly shapes, even when there is a low penalty.

Since there is not a well defined minimum for the CV score, we opt for presenting two different cases.
One is more conservative, in the sense that it has a stronger penalty that allows only small deviations from the concordance power-law model.
For this one we choose $\alpha_p=1$.

The other leaves more freedom to the data, as we choose a more relaxed penalty $\alpha_p=0.01$.
A reconstruction with $\alpha_p \ll 0.01$ is pretty much uninformative. In fact recall that  the free parameters in our MCMC runs are the physical baryon density $\omega_{b }$, the physical cold dark matter density $\omega_{\rm{cdm}}$, the rescaled Hubble parameter $h$, the optical depth to reionization $\tau_{\rm{reio}}$, and the value of the five knots of the spline that we used to parametrize the shape of the PPS. At such low penalty values the reconstruction transfers in part the features of the radiation transfer function and the effect of the optical depth to reionization into the PPS opening up degeneracies in parameter space.

\section{Results}
\label{sec:results}
Here we present the results with the latest \textit{Planck} likelihood (2015 release) and all the large scale structure power spectrum data (\textit{Planck} Lensing 2015, WiggleZ, CFHTLenS, and SDSS DR7), with  the two different roughness penalties ($\alpha_p= 1$ and  $\alpha_p=0.01$) justified above.

As discussed in refs. \cite{Planck:CosmologicalParameters, 2015MNRAS.451.2877M, 2016arXiv160105786J, 2016MNRAS.459..971K,2015arXiv151203626C} there is a tension between the inferred matter power spectrum amplitude from  CMB and from CFHTLenS,  which may arise from possible systematic errors in the photometric redshifts of CFHTLens. For this reason we present results first without and then  with CFHTLens.

\subsection{Reconstruction without CFHTLens}
\label{sec:resnoCFHT}
\begin{figure}[t]
	\begin{subfigure}{0.5\textwidth}
		\centering
		\includegraphics[width=\textwidth]{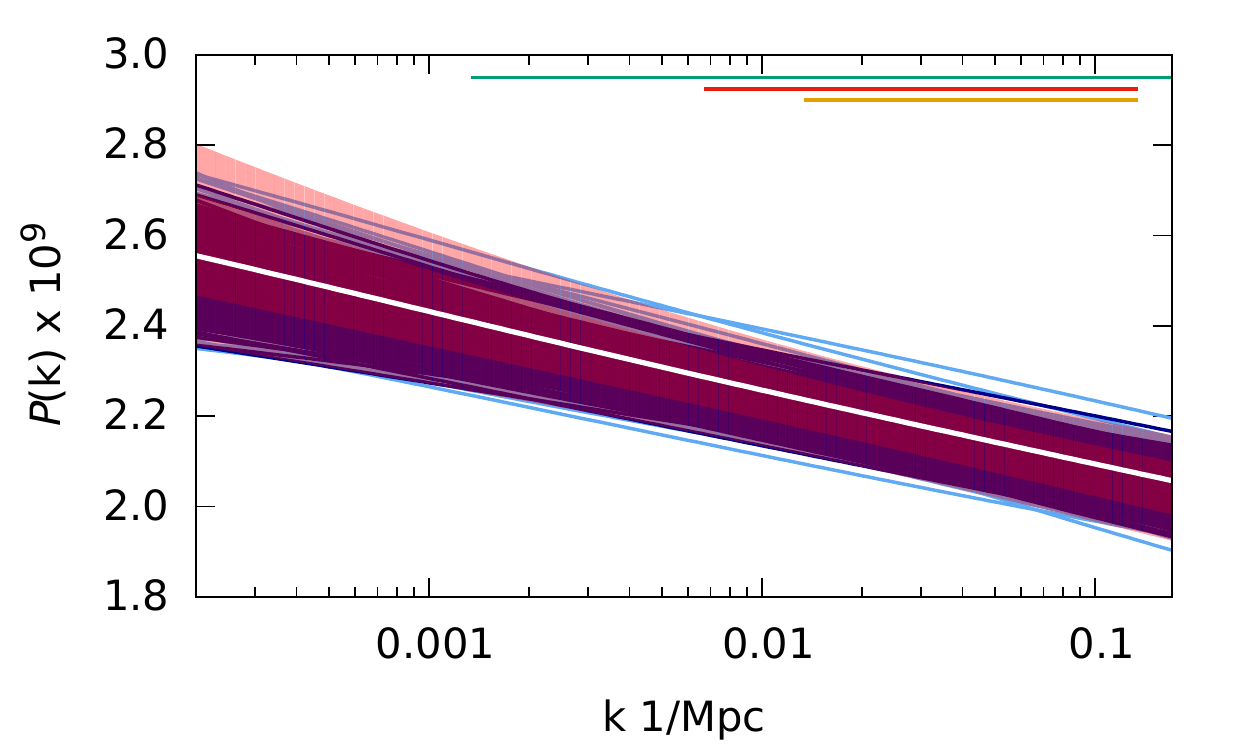}
		\caption{$\alpha_p=1$.}
        \label{fig:SWP_a1_PPS}
	\end{subfigure}
	\begin{subfigure}{0.5\textwidth}
  		\centering
		\includegraphics[width=\textwidth]{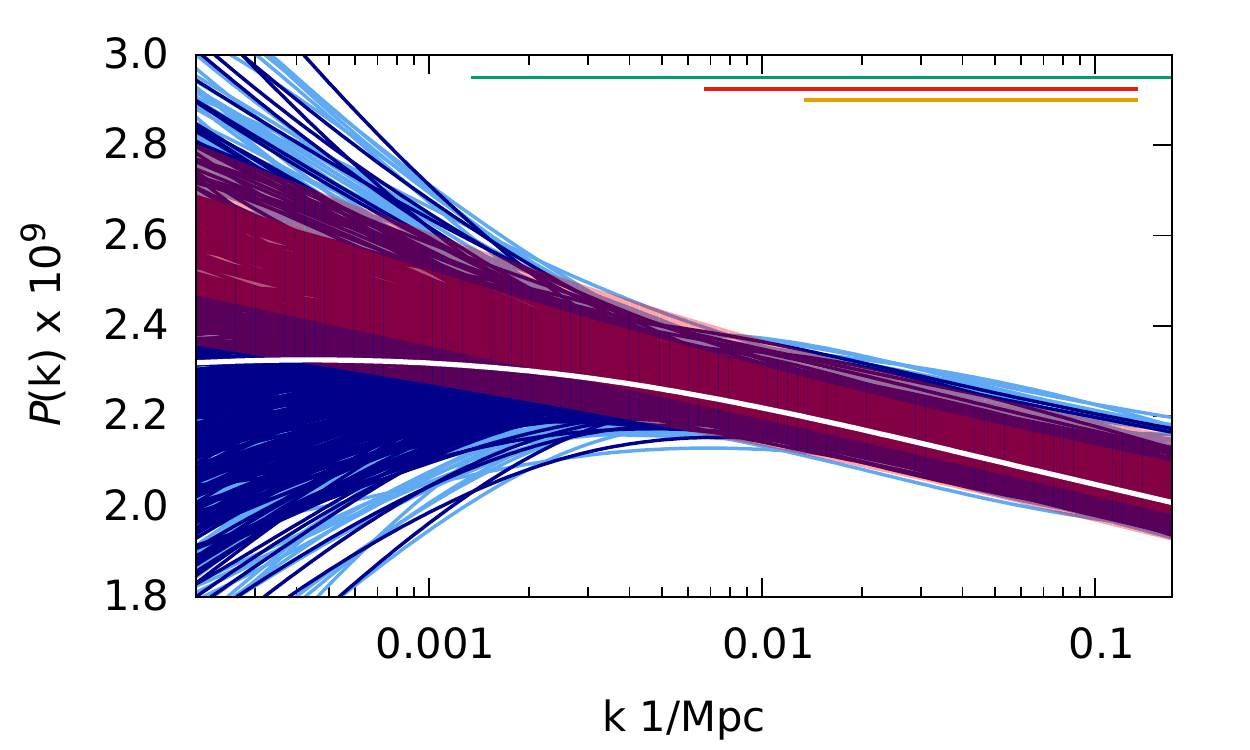}
		\caption{$\alpha_p=0.01$.}
        \label{fig:SWP_a001_PPS}
	\end{subfigure}
\caption{Reconstructed PPS. The  best fit  reconstruction is shown  in white. Errors are shown by plotting in dark blue (light blue) 400 spline picked at random among the 68.27\% most likely points (points in the range 68.27\% - 95.45\%) in the MCMC. The red (pale red) region shows the 68\% (95\%) confidence intervals for \textit{Planck} 2015 TT, TE, EE + Low P. The colour-bars on the upper side show the scales probed by each experiment as in figure \ref{fig:CoveredScales}, green for PlanckLens, red for WiggleZ, gold for SDSS DR7. PlanckCMB15 covers the whole plot.}
\label{fig:SWP_PPS}
\end{figure}

In figure \ref{fig:SWP_a1_PPS} and \ref{fig:SWP_a001_PPS} we show the reconstructed PPS  for $\alpha_p=1$ and $\alpha_p=0.01$ respectively.  The colour-bars on the upper side show the scales probed by each experiment as in figure \ref{fig:CoveredScales}, green for PlanckLens, red for WiggleZ, gold for SDSS DR7. PlanckCMB15 covers the whole plot. The best fit  reconstruction is shown in yellow and  errors are shown by plotting in dark blue (light blue)  a random sample of 400  reconstructions  chosen among the 68.27\% most likely points (points in the range 68.27\% - 95.45\%) in the MCMC. The  95.5\% confidence regions appear to coincide with the 68.3\%: this is because the reconstructed spectra are simply more wiggly and are not allowed to deviate more, and consistently across scales, from the best fit.

In the figure the red and pale red regions show the 68 and 95\% confidence intervals for the standard power law $\Lambda$CDM \textit{Planck} 2015 TT, TE, EE + Low P analysis \cite{Planck:CosmologicalParameters2015}.

Note that for the  more conservative choice of the penalty, errors of the reconstructed PPS are comparable with errors from \textit{Planck} parametric fit at all scales. For the less conservative penalty this is also true on scales corresponding to $\ell >30$. This did not happen with the previous generation of cosmological data (see \cite{Peiris:bPlanck2010}) where the reconstructed PPS was significantly less constrained than with a power law fit.

The additional freedom in the PPS allowed by the lower penalty $\alpha_p = 0.01$ is used on scales corresponding to low CMB multipoles $\ell<30$. These scales are dominated by cosmic variance and are known to be lower than the standard $\Lambda$CDM prediction e.g., \cite{Planck:CMBPowerSpectra,2015arXiv151007929S,2015MNRAS.451.2978C, 2010AdAst2010E..92C} and refs therein.

\begin{figure}[tbp]
	\begin{subfigure}{0.5\textwidth}
		\centering
		\includegraphics[width=\textwidth]{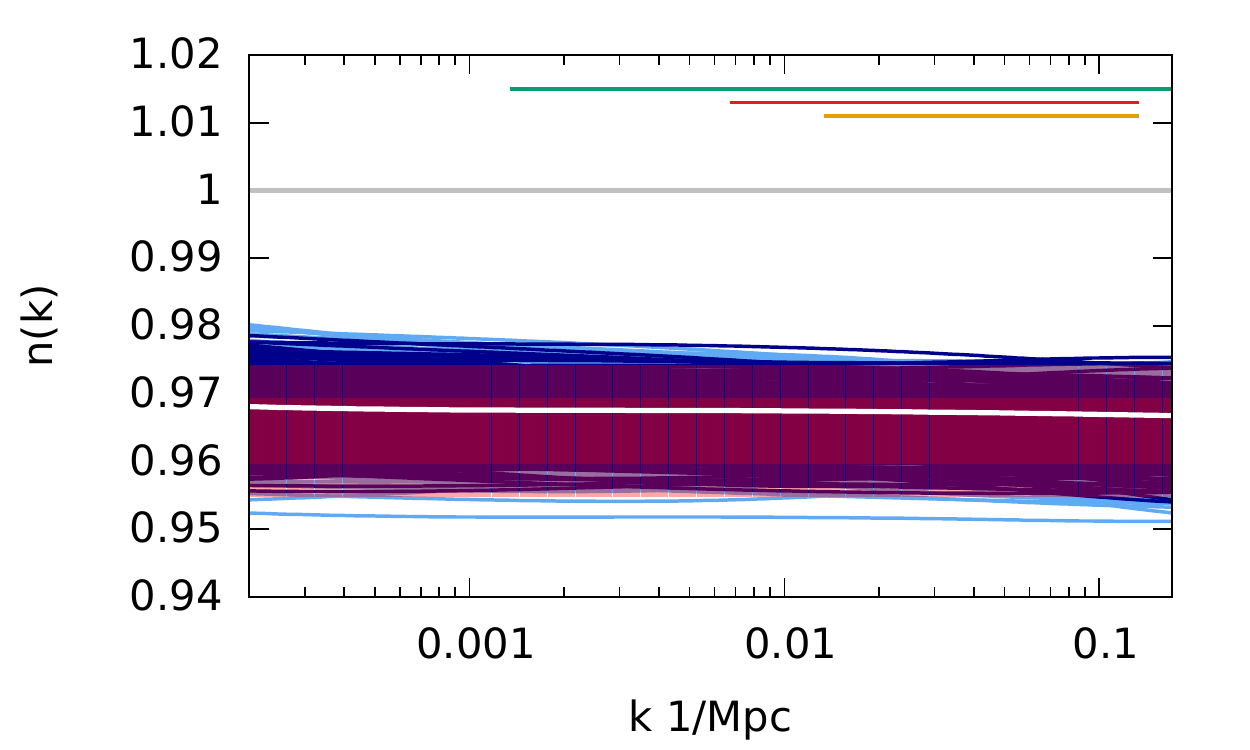}
		\caption{$\alpha_p=1$.}
        \label{fig:SWP_a1_nk}
	\end{subfigure}
	\begin{subfigure}{0.5\textwidth}
  		\centering
		\includegraphics[width=\textwidth]{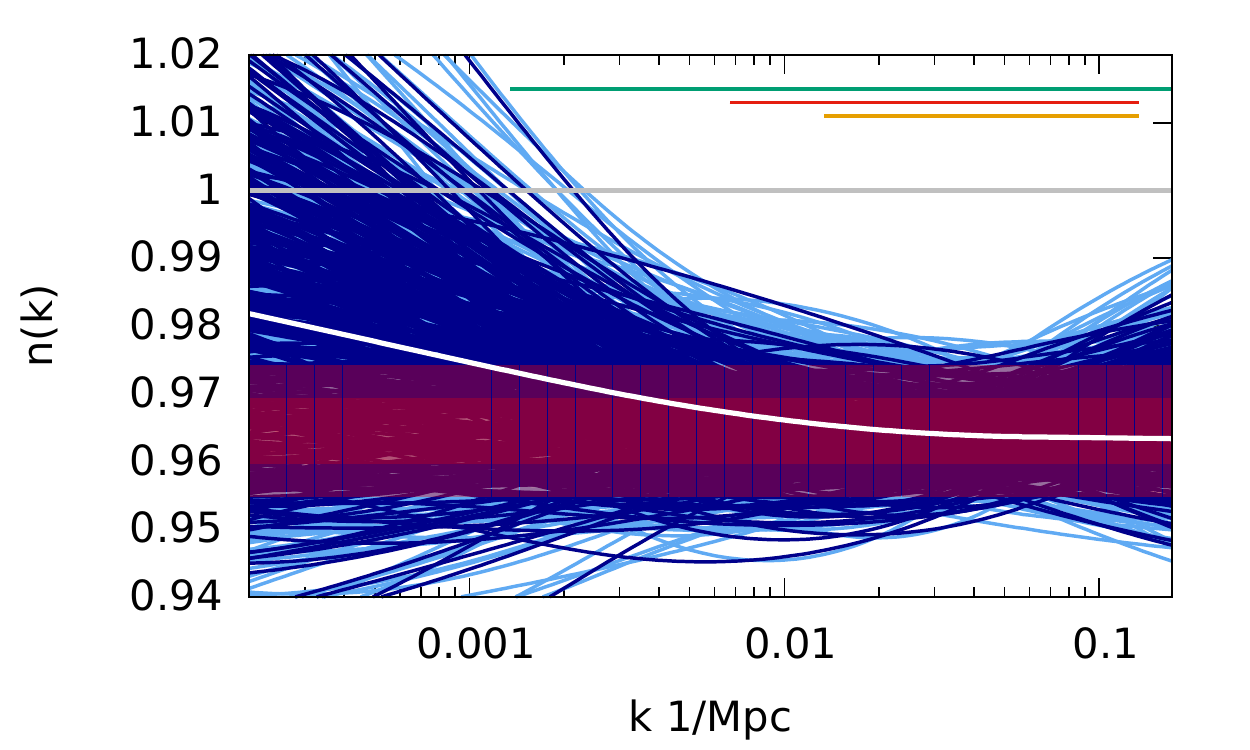}
		\caption{$\alpha_p=0.01$.}
        \label{fig:SWP_a001_nk}
	\end{subfigure}
\caption{Power spectrum spectral index of the reconstructed PPSs. The  white line corresponds to the best fit reconstruction. Errors are shown by plotting in dark blue (light blue) 400 reconstructions randomly selected from the 68.27\% most likely points (points in the range 68.27\% - 95.45\%) in the MCMC. The red (pale red) region shows 68\% (95\%) confidence intervals for the power law \textit{Planck} 2015 TT, TE, EE + Low P fit. The colour-bars on the upper side show the scales probed by each experiment as in figure \ref{fig:CoveredScales}, green for PlanckLens red for WiggleZ, gold for SDSS DR7. PlanckCMB15 covers the whole plot. In the right figure, the grey line is $n(k) \equiv 1$, i.e., scale invariance.}
\label{fig:SWP_nk}
\end{figure}
In figure \ref{fig:SWP_a1_nk} and  \ref{fig:SWP_a001_nk} we also  show the reconstructed $n(k)\equiv d\ln P(k)/d\ln k$ ($\alpha_p=1$ and  $\alpha_p=0.01$) for ease of comparison with the standard power law results.\footnote{Recall that  the quantity that was actually reconstructed using cross-validation to find the optimal penalty is  in reality the power spectrum.}
We find no evidence that any scale dependence of  the power spectrum spectral slope is necessary, which is in agreement with previous analyses. However with  this new data set we  find that $n=1$ is highly disfavoured by the data, in particular for $\alpha_p=1$ the  significance of the departure from scale invariance is comparable with that obtained when adopting the ``inflation--motivated'' power-law prior. Even for the more flexible reconstruction,  not even one point of the more than $4\times 10^5$ MCMC points falls near scale invariance.

The results shown in Figs.~\ref{fig:SWP_PPS} and \ref{fig:SWP_nk} offer a powerful confirmation of the inflationary paradigm, justify the adoption  of the  inflationary prior in cosmological analyses.

\begin{figure}[tbp]
	\begin{subfigure}{0.5\textwidth}
		\centering
		\includegraphics[width=\textwidth]{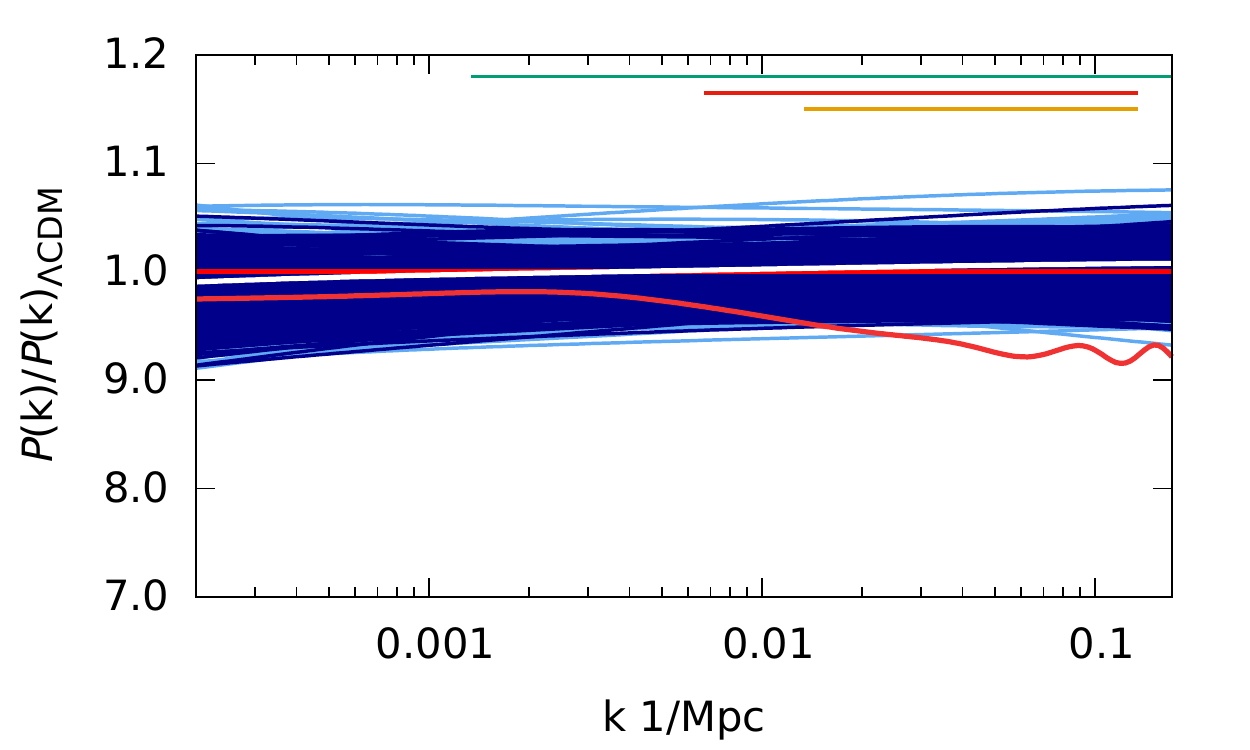}
		\caption{$\alpha_p=1$.}
        \label{fig:SWP_a1_ratio}
	\end{subfigure}
	\begin{subfigure}{0.5\textwidth}
  		\centering
		\includegraphics[width=\textwidth]{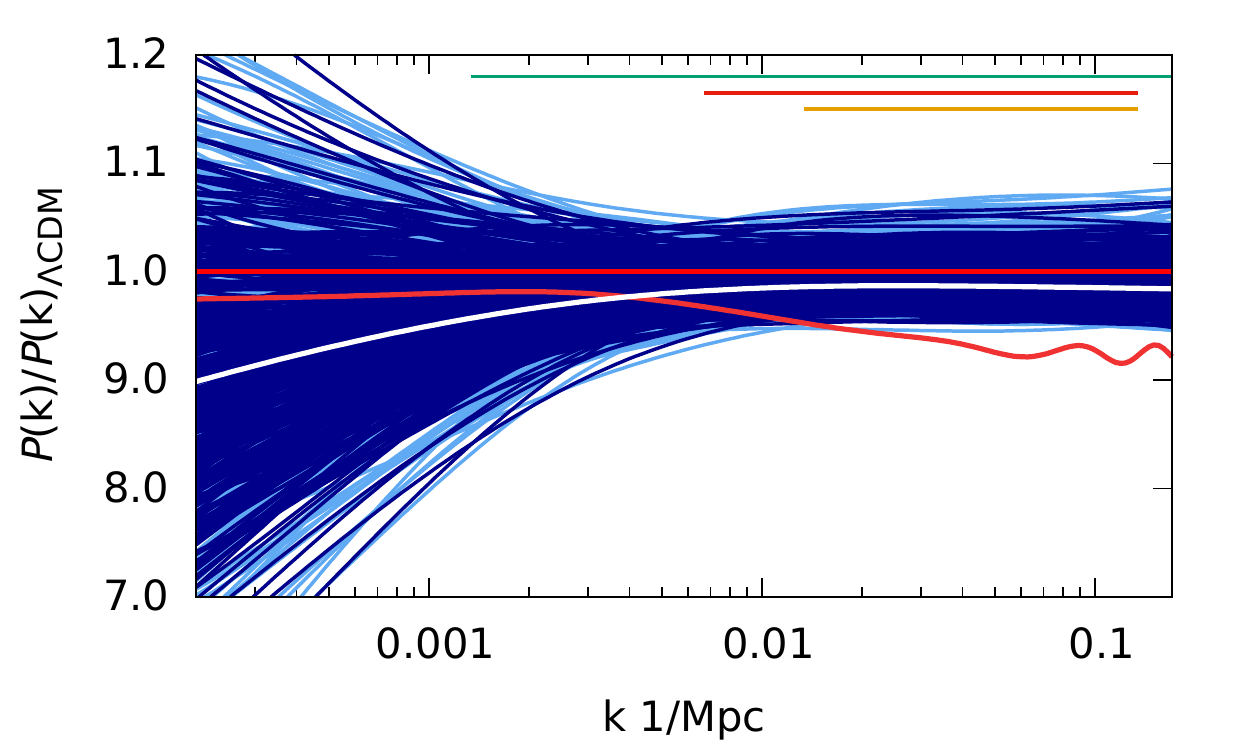}
		\caption{$\alpha_p=0.01$.}
        \label{fig:SWP_a001_ratio}
	\end{subfigure}
\caption{Reconstructed PPS divided by the \textit{Planck} 2015 TT, TE, EE + LowP + Lensing power-law PPS best-fit using the same conventions as Figs.\ref{fig:SWP_PPS}  for the legend and the reconstructed $P(k)$. The red lines show the small-scales power suppression effect due to massive neutrinos.
 The upper line is  the  $\Sigma m_\nu = 0 $ eV theoretical prediction based on the conditional best fit to \textit{Planck} 2015 TT, TE, EE + Low P + Lensing + BAO + JLA + $H_0$ data, the lower line is the same with $\Sigma m_\nu = 0.2$ eV.}
        \label{fig:SWP_ratio}
\end{figure}

Finally in figs. \ref{fig:SWP_a1_ratio} and \ref{fig:SWP_a001_ratio}  we show the ratio of the reconstructed PPS to the best fit  \textit{Planck} 2015 (temperature, polarisation, and lensing) power law model. 

The reconstruction is fully compatible with the parametric fit. The figure also shows the  expected effect of small scale power suppression due to massive neutrino free-streaming for two representative values of neutrino masses $\Sigma m_\nu =0$ eV and $0.2$ eV.
The two models are the conditional (i.e., keeping $\Sigma m_\nu$ fixed at the required value) best fit to the data (\textit{Planck} 2015 TT, TE, EE + Low P + Lensing + BAO + JLA + $H_0$ data).
Clearly models with $\Sigma m_\nu > 0.2$ eV are highly disfavoured by the data even with this minimally parametric reconstruction: not a single step of a $4\times 10^5$ size MCMC goes near the  $\Sigma m_\nu = 0.2$ eV line.
This of course does not exclude the --- admittedly contrived --- case with a arbitrarily large neutrino mass inducing a small scale power suppression which is cancelled by a compensating boost of the PPS on the same scales. Occam's razor disfavours this scenario.

\subsection{Reconstruction with CFHTLens}
\label{sec:resCFHT}
\begin{figure}[tbp]
	\begin{subfigure}{0.5\textwidth}
		\centering
		\includegraphics[width=\textwidth]{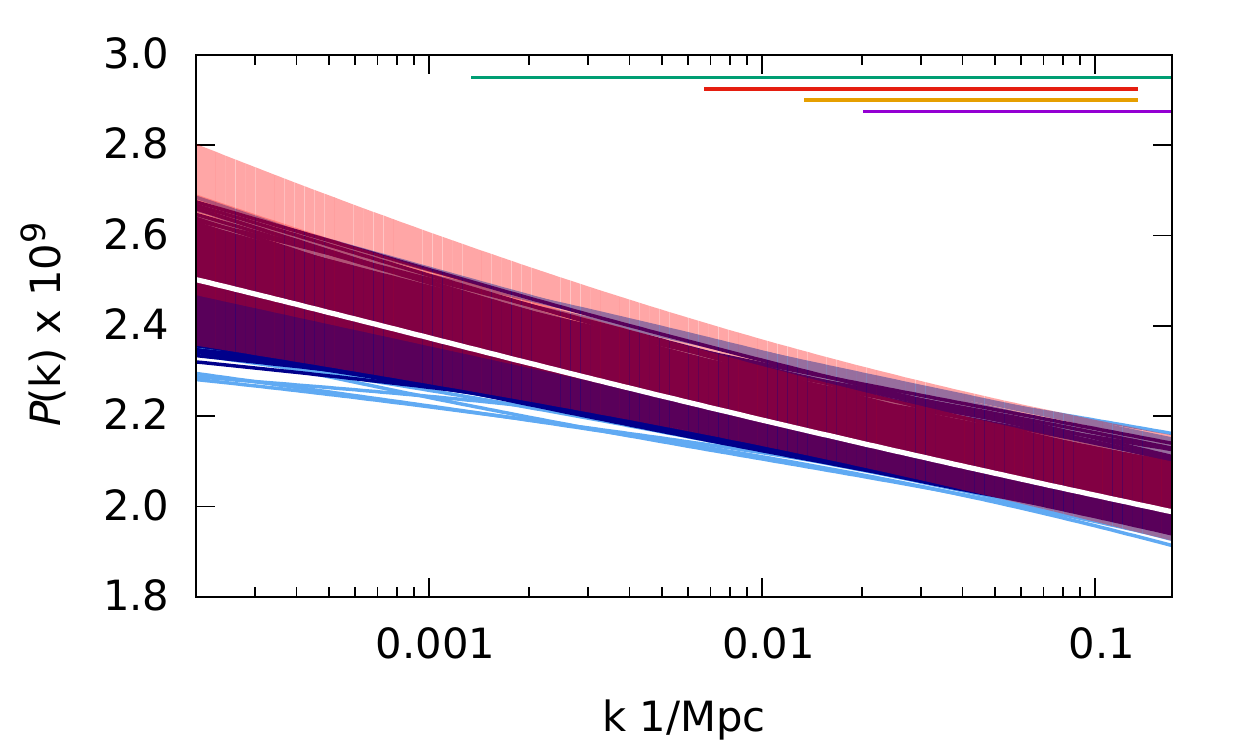}
		\caption{$\alpha_p=1$.}
		\label{fig:4expA1-PPS+Planck}
	\end{subfigure}
	\begin{subfigure}{0.5\textwidth}
  		\centering
		\includegraphics[width=\textwidth]{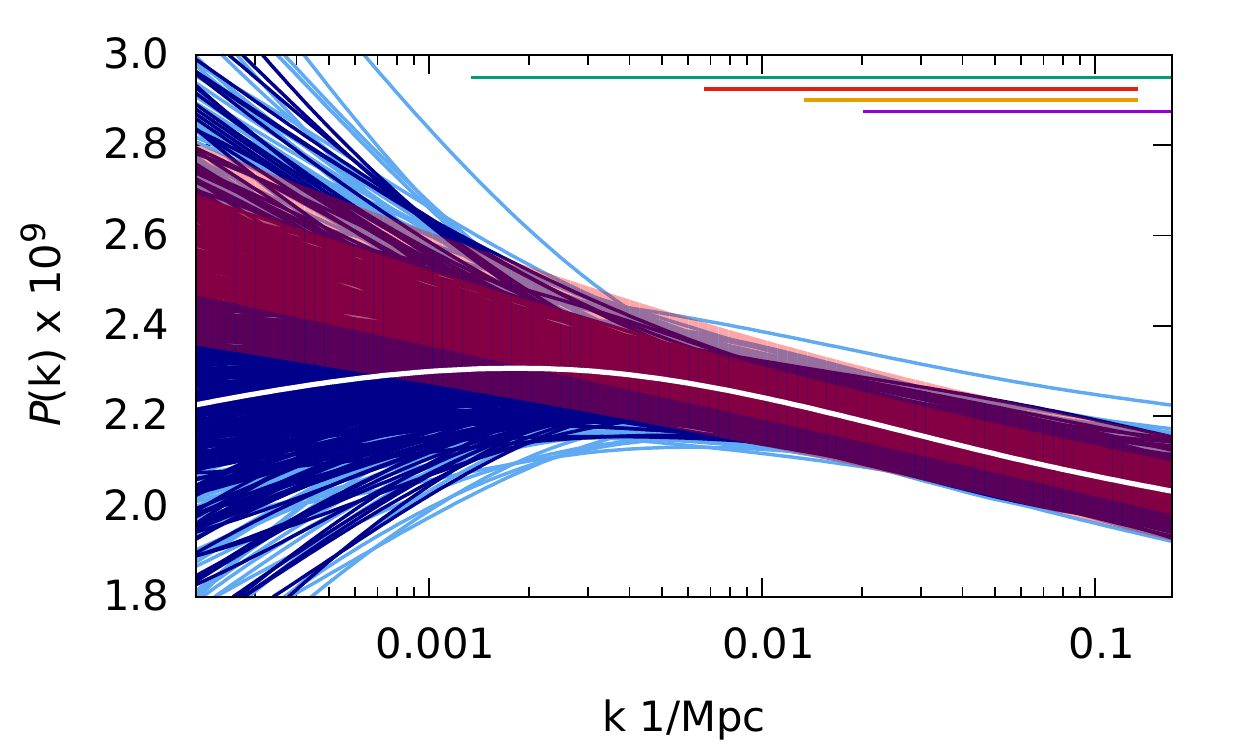}
		\caption{$\alpha_p=0.01$.}
		\label{fig:4expA001-PPS+Planck}
	\end{subfigure}

\caption{Reconstructed PPS. Refer to figure \ref{fig:SWP_PPS} for explanation and colour code. In addition, the purple line shows the scales covered by CFHTLenS.}
\label{fig:SWCP_PPS}
\end{figure}

\begin{figure}[tbp]
	\begin{subfigure}{0.5\textwidth}
		\centering
		\includegraphics[width=\textwidth]{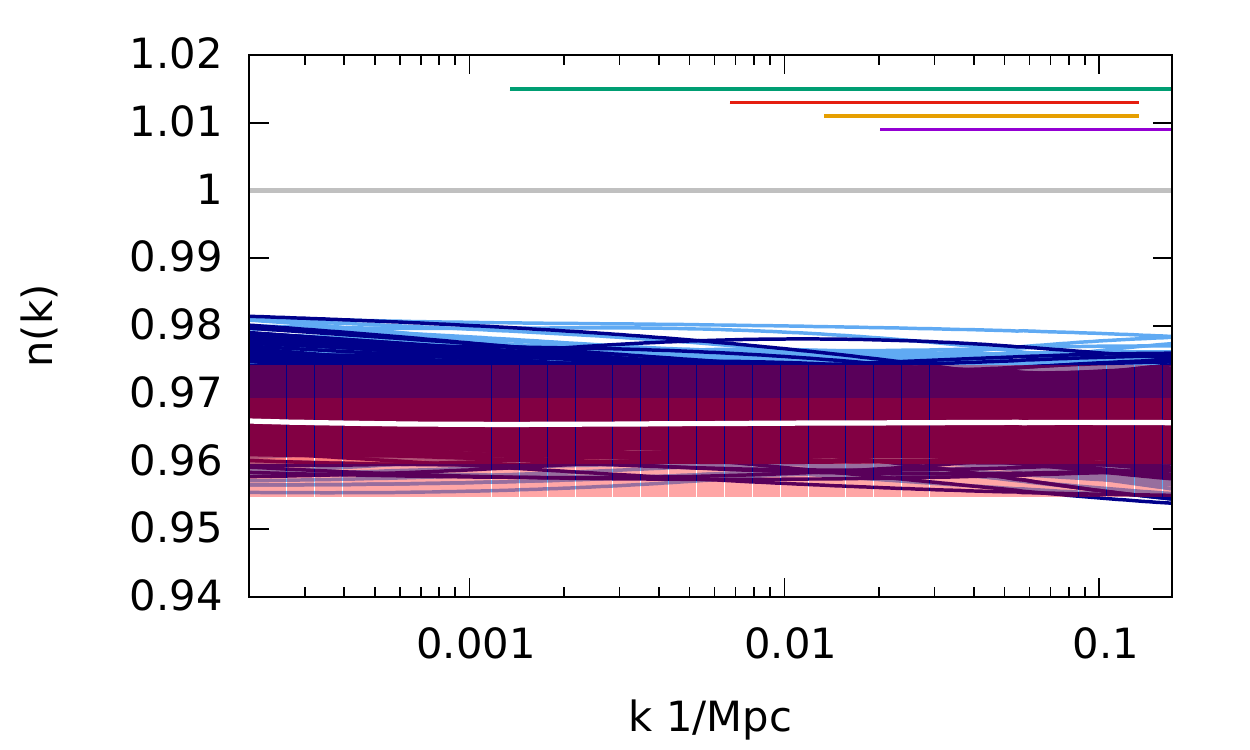}
		\caption{$\alpha_p=1$.}
		\label{fig:4expA1-n+Planck}
	\end{subfigure}
	\begin{subfigure}{0.5\textwidth}
  		\centering
		\includegraphics[width=\textwidth]{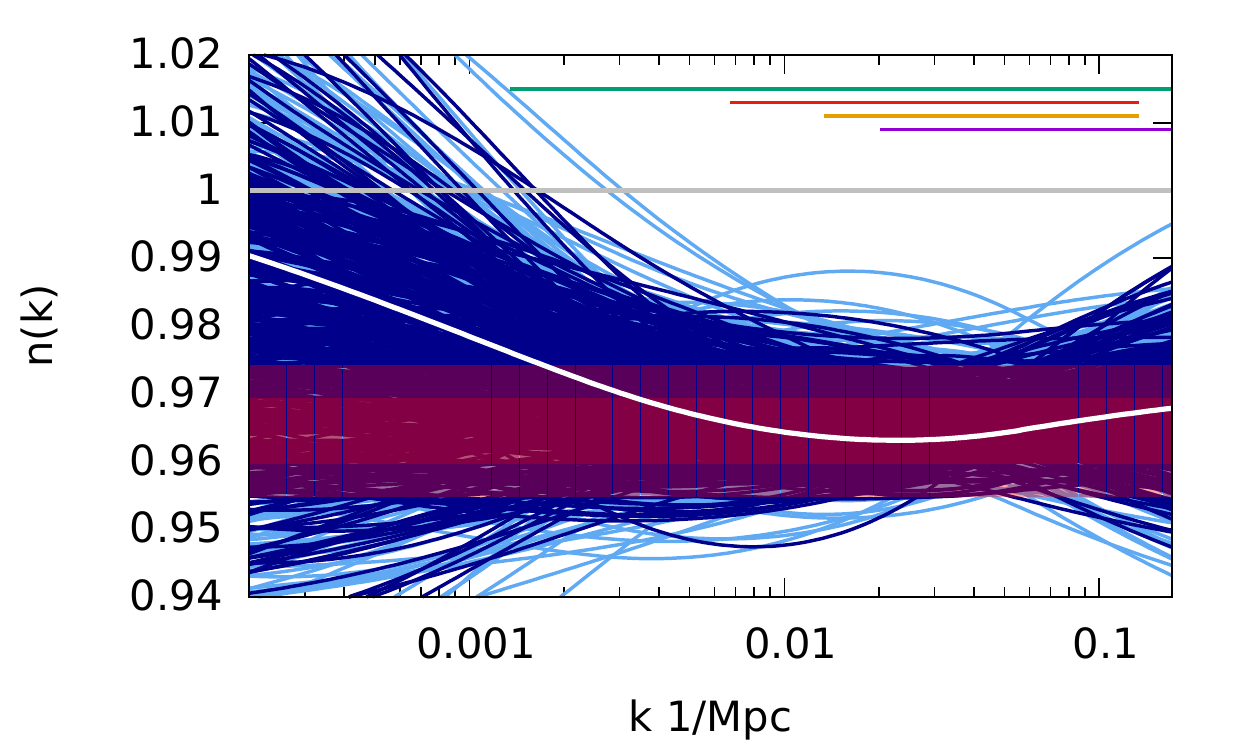}
		\caption{$\alpha_p=0.01$.}
		\label{fig:4expA001-n+Planck}
	\end{subfigure}
\caption{Power spectrum spectral index of the reconstructed PPSs. Refer to figure \ref{fig:SWP_nk} for explanation and colour code. In addition, the purple line shows the scales covered by CFHTLenS.}
\label{fig:SWCP_nk}
\end{figure}

\begin{figure}[tbp]
	\begin{subfigure}{0.5\textwidth}
		\centering
		\includegraphics[width=\textwidth]{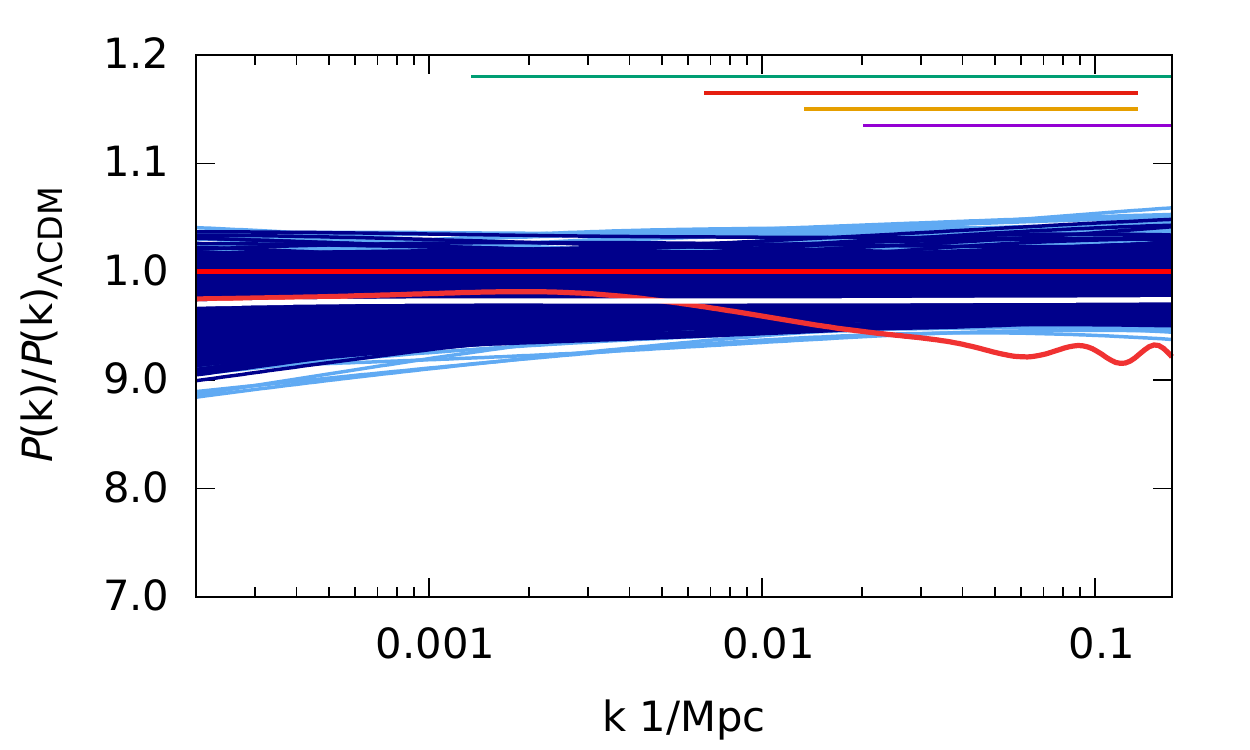}
		\caption{$\alpha_p=1$.}
		\label{fig:4expA1-Ratio+Nu}
	\end{subfigure}
	\begin{subfigure}{0.5\textwidth}
  		\centering
		\includegraphics[width=\textwidth]{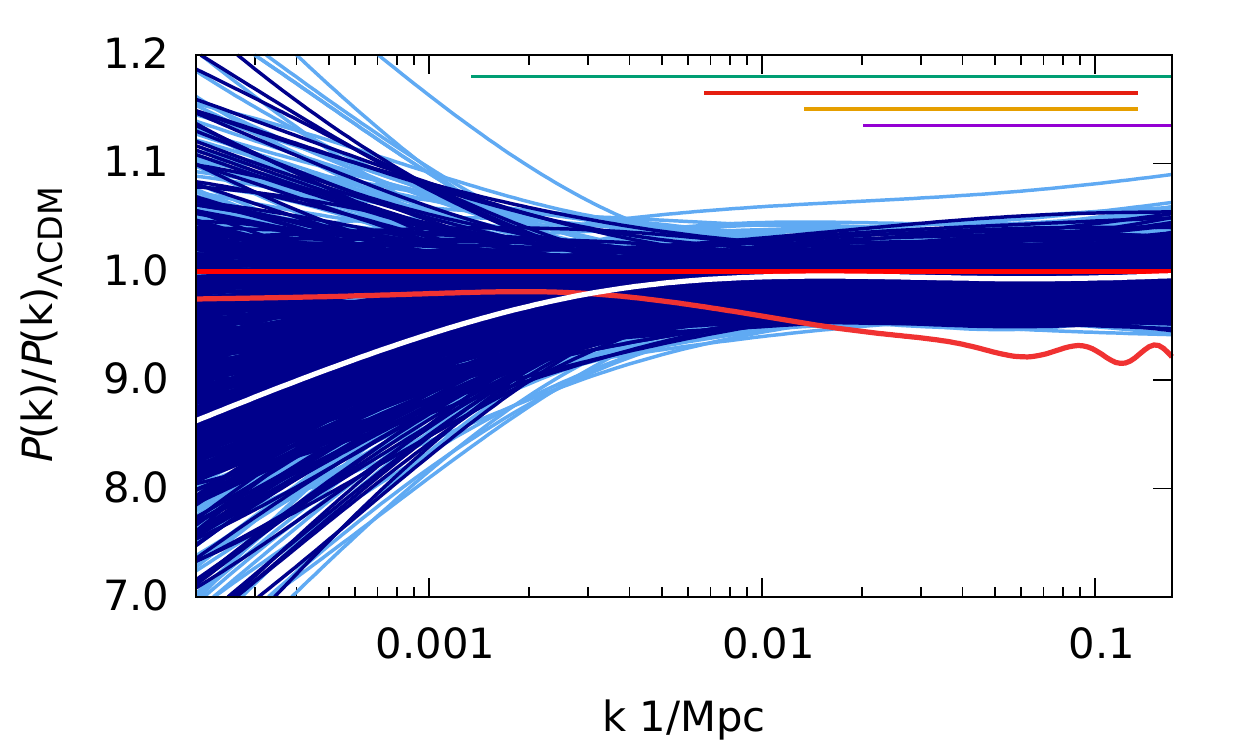}
		\caption{$\alpha_p=0.01$.}
		\label{fig:4expA001-Ratio+Nu}
	\end{subfigure}
\caption{Reconstructed PPS  relative to  \textit{Planck} 2015 TT, TE, EE + LowP + Lensing power-law PPS best-fit. Refer to figure \ref{fig:SWP_ratio} for explanation and colour code. In addition, the purple line shows the scales covered by CFHTLenS. }
\label{fig:SWCP_ratio}
\end{figure}

The reconstructed $P(k)$, $n(k)$ and $P(k)$ relative to the power law best fit  are shown in figures~\ref{fig:SWCP_PPS}, \ref{fig:SWCP_nk}, and \ref{fig:SWCP_ratio} using the same conventions as in figures~\ref{fig:SWP_PPS}, \ref{fig:SWP_nk}, and \ref{fig:SWP_ratio}.


Comparison with the results of section \ref{sec:resnoCFHT} (in figures~\ref{fig:SWP_PPS}, \ref{fig:SWP_nk}, \ref{fig:SWP_ratio}) shows that qualitatively the reconstructions are very similar, there is no strong evidence for deviations from the power law behaviour and scale invariance is still excluded. However quantitatively some differences may be appreciated.  Adding the CFHTLenS datasets has the effect of lowering the overall PPS normalisation (clearly visible by comparison with figure \ref{fig:SWP_PPS}).

The ratio with Planck power law best-fit in figure~\ref{fig:SWCP_ratio} highlights how, independently from our choice of datasets, high neutrino masses are disfavoured.
Quantitatively the $\Sigma m_\nu > 0.2$ eV bound is excluded at more than 95\% confidence if we assume a power law PPS, as discussed in section~\ref{sec:resnoCFHT}.

For completeness we also report the recovered values  and errors for all the model parameters in  table~\ref{tab:FinalResults1} and table~\ref{tab:FinalResults001}  for the two penalties $\alpha_p=1$ and $\alpha_p=0.01$ respectively.

\begin{table}[tbp]
\centering
\begin{tabular}{|lcccc|}
\hline
\multicolumn{5}{|c|}{Without CFHTLenS $\qquad -\ln{\cal L}_\mathrm{min} =6727.02$} \\
Param & best-fit & mean$\pm\sigma$ & 95\% lower & 95\% upper \\
\hline 
$100~\omega_{b }$ &$2.226$ & $2.229_{-0.015}^{+0.015}$ & $2.199$ & $2.258$ \\ 
$\omega_{\rm{cdm}}$ &$0.1194$ & $0.1191_{-0.0011}^{+0.0011}$ & $0.1168$ & $0.1214$ \\ 
$h$ &$0.6795$ & $0.6815_{-0.0053}^{+0.0052}$ & $0.6712$ & $0.6919$ \\ 
$\tau$ &$0.0542$ & $0.05982_{-0.013}^{+0.0092}$ & $0.04001$ & $0.07955$ \\ 
$10^{+9}K_{1 }$ &$2.788$ & $2.736_{-0.12}^{+0.11}$ & $2.507$ & $2.965$ \\ 
$10^{+9}K_{2 }$ &$2.546$ & $2.534_{-0.07}^{+0.07}$ & $2.395$ & $2.673$ \\ 
$10^{+9}K_{3 }$ &$2.307$ & $2.32_{-0.049}^{+0.043}$ & $2.231$ & $2.412$ \\ 
$10^{+9}K_{4 }$ &$2.072$ & $2.108_{-0.053}^{+0.036}$ & $2.028$ & $2.194$ \\ 
$10^{+9}K_{5 }$ &$1.872$ & $1.9_{-0.059}^{+0.047}$ & $1.802$ & $2.006$ \\  
\hline
\multicolumn{5}{|c|}{With CFHTLenS $\qquad -\ln{\cal L}_\mathrm{min} =6777.64$} \\
Param & best-fit & mean$\pm\sigma$ & 95\% lower & 95\% upper \\
\hline 
$100~\omega_{b }$ &$2.244$ & $2.236_{-0.014}^{+0.015}$ & $2.207$ & $2.264$ \\ 
$\omega_{\rm{cdm}}$ &$0.1182$ & $0.1182_{-0.001}^{+0.0011}$ & $0.1161$ & $0.1204$ \\ 
$h$ &$0.6867$ & $0.6854_{-0.0051}^{+0.0048}$ & $0.6757$ & $0.6953$ \\ 
$\tau$ &$0.05239$ & $0.05852_{-0.013}^{+0.0087}$ & $0.04001$ & $0.07795$ \\ 
$10^{+9}K_{1 }$ &$2.721$ & $2.697_{-0.12}^{+0.12}$ & $2.463$ & $2.931$ \\ 
$10^{+9}K_{2 }$ &$2.512$ & $2.504_{-0.071}^{+0.069}$ & $2.363$ & $2.645$ \\ 
$10^{+9}K_{3 }$ &$2.281$ & $2.3_{-0.048}^{+0.041}$ & $2.213$ & $2.389$ \\ 
$10^{+9}K_{4 }$ &$2.065$ & $2.098_{-0.051}^{+0.035}$ & $2.021$ & $2.182$ \\ 
$10^{+9}K_{5 }$ &$1.869$ & $1.9_{-0.058}^{+0.045}$ & $1.802$ & $2.003$ \\
\hline 
\end{tabular} \\ 
\caption{Best fit, mean and confidence intervals for the MCMC parameters in the reconstruction with $\alpha_p=1$}
\label{tab:FinalResults1}
\end{table}

\begin{figure}[tbp]
    \begin{subfigure}{0.5\textwidth}
		\includegraphics[width=\textwidth]{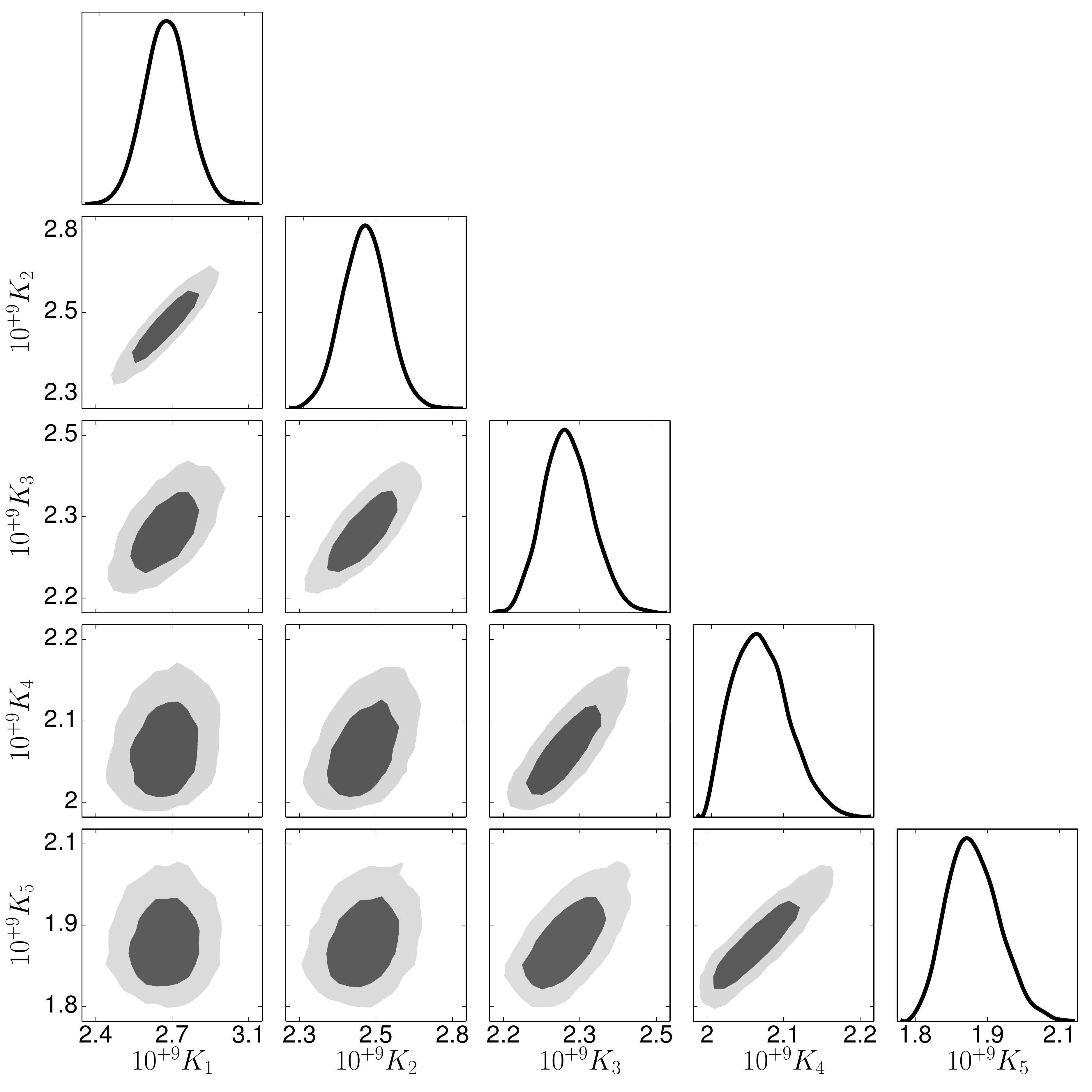}
		\caption{$\alpha_p=1$.}
		\label{fig:SWCP_a1_triangle_knots}
	\end{subfigure}\hfill%
    \begin{subfigure}{0.5\textwidth}
		\includegraphics[width=\textwidth]{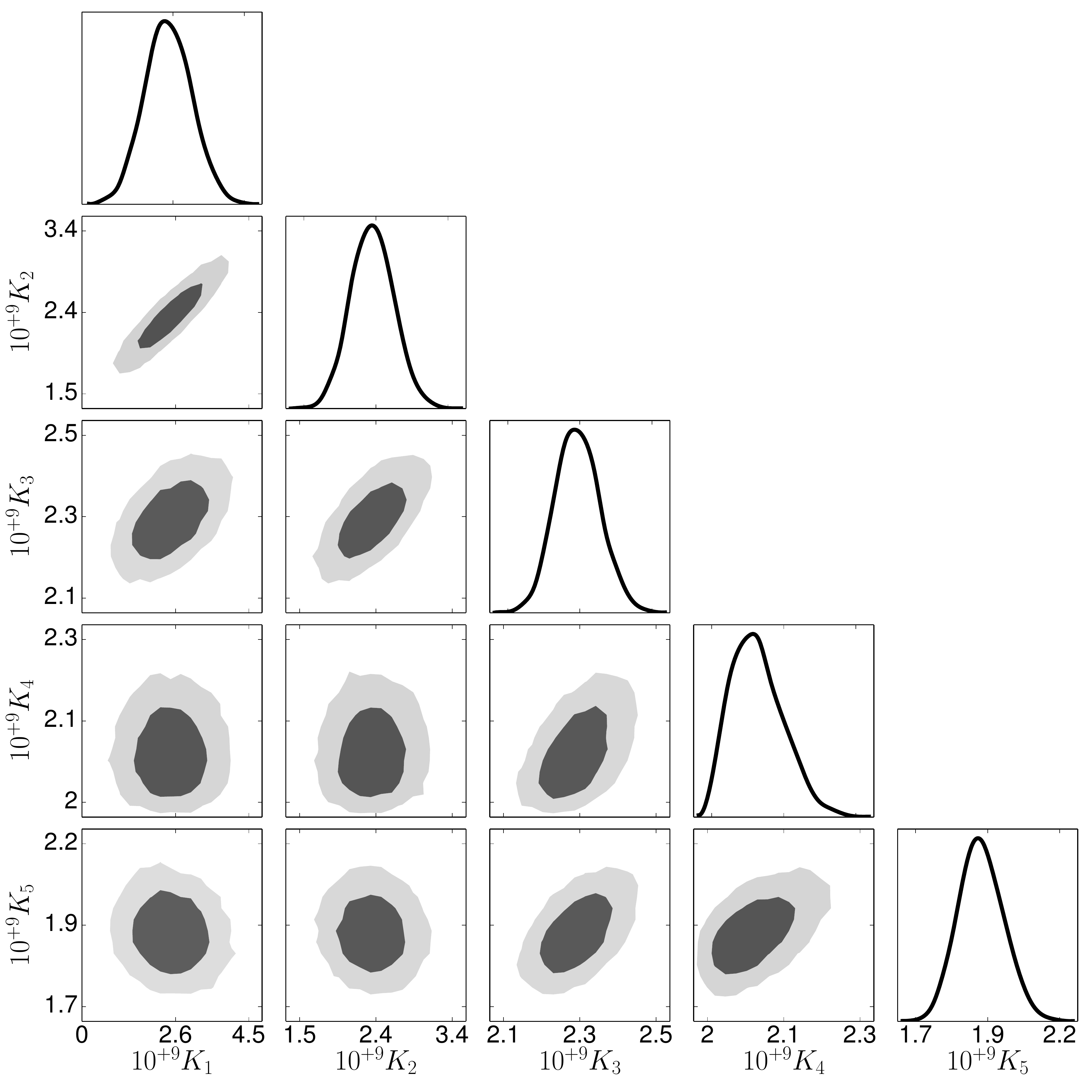}
		\caption{$\alpha_p=0.01$.}
		\label{fig:SWCP_a001_triangle_knots}
	\end{subfigure}
\newline
    \begin{subfigure}{0.5\textwidth}
        \centering
		\includegraphics[width=0.7\textwidth]{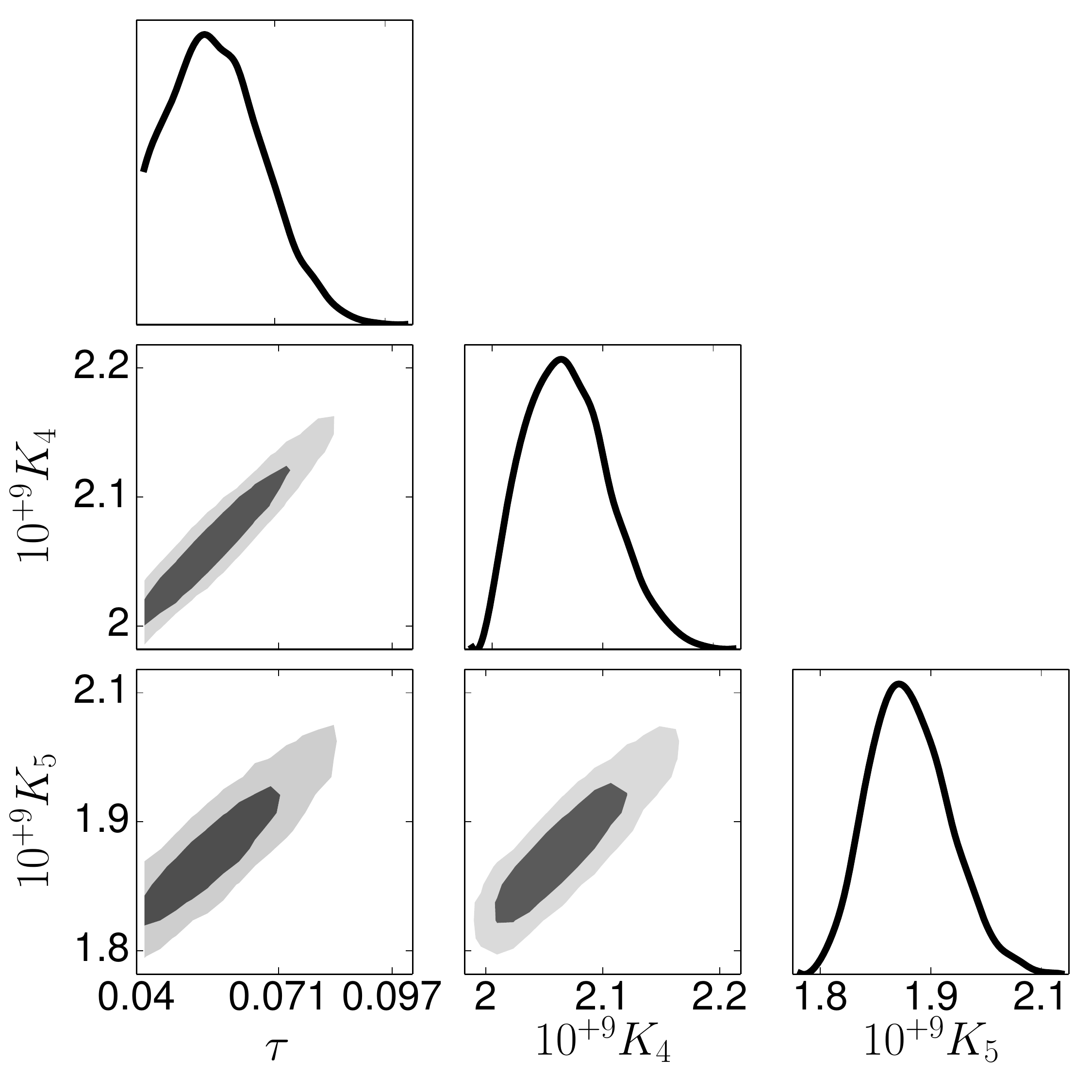}
		\caption{$\alpha_p=1$.}
		\label{fig:SWCP_a1_triangle_tau}
	\end{subfigure}\hfill%
    \begin{subfigure}{0.5\textwidth}
        \centering
		\includegraphics[width=0.7\textwidth]{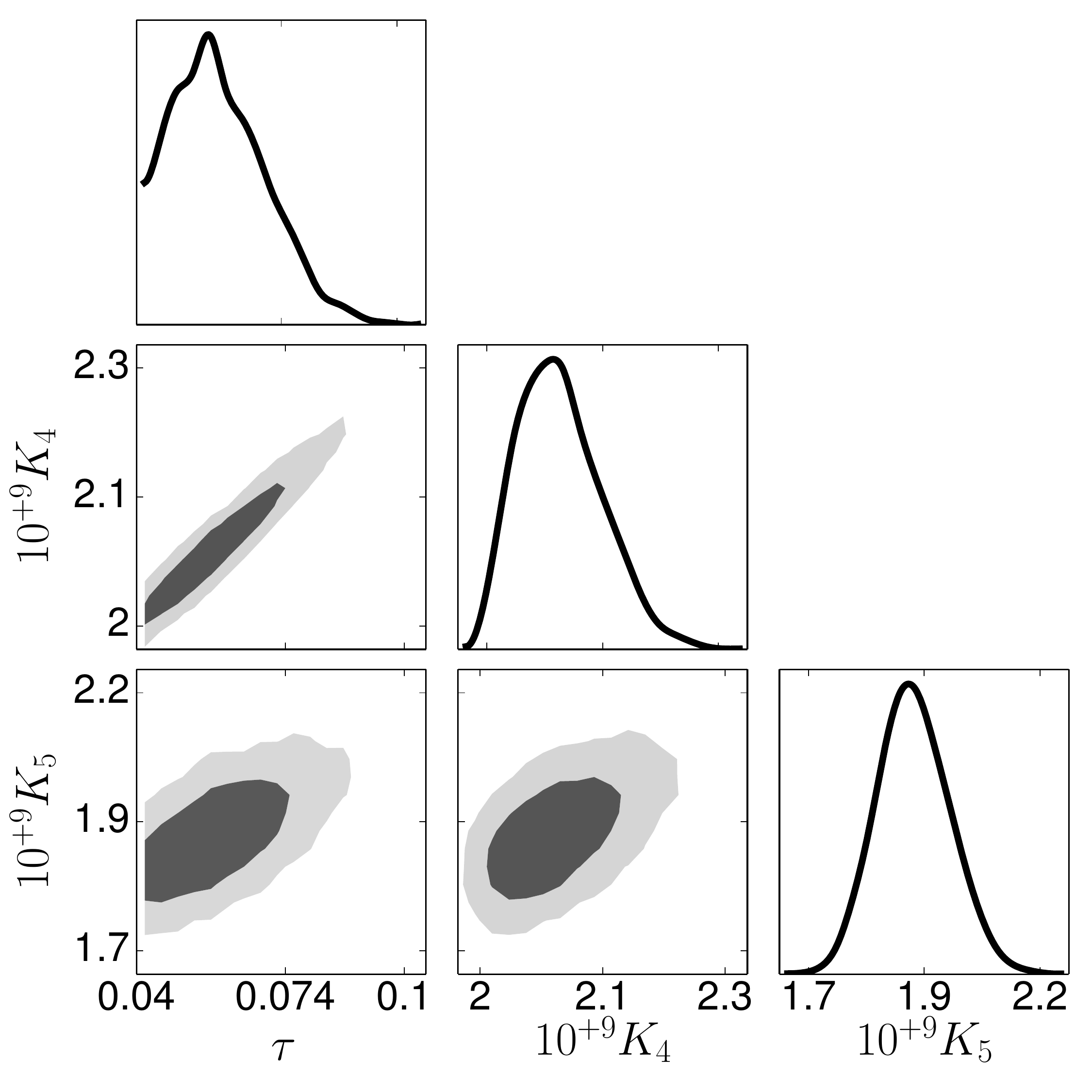}
		\caption{$\alpha_p=0.01$.}
		\label{fig:SWCP_a001_triangle_tau}
	\end{subfigure}
\caption{Triangular plots for the run with all the datasets combined. We refer to the value of the spline function evaluated at the $i$-th knot as $K_i$.}
\end{figure}

\begin{table}[!tbp]
\centering
\begin{tabular}{|lcccc|} 
\hline
\multicolumn{5}{|c|}{Without CFHTLenS $\qquad -\ln{\cal L}_\mathrm{min} =6726.32$} \\
Param & best-fit & mean$\pm\sigma$ & 95\% lower & 95\% upper \\
\hline 
$100~\omega_{b }$ &$2.239$ & $2.226_{-0.016}^{+0.016}$ & $2.194$ & $2.259$ \\ 
$\omega_{\rm{cdm}}$ &$0.1173$ & $0.1193_{-0.0012}^{+0.0013}$ & $0.1168$ & $0.1218$ \\ 
$h$ &$0.6898$ & $0.6806_{-0.006}^{+0.0055}$ & $0.6694$ & $0.6922$ \\ 
$\tau$ &$0.07136$ & $0.05974_{-0.014}^{+0.0088}$ & $0.04$ & $0.08061$ \\ 
$10^{+9}K_{1 }$ &$1.968$ & $2.351_{-0.69}^{+0.67}$ & $1.014$ & $3.711$ \\ 
$10^{+9}K_{2 }$ &$2.159$ & $2.351_{-0.3}^{+0.29}$ & $1.769$ & $2.934$ \\ 
$10^{+9}K_{3 }$ &$2.272$ & $2.31_{-0.062}^{+0.059}$ & $2.191$ & $2.431$ \\ 
$10^{+9}K_{4 }$ &$2.148$ & $2.106_{-0.056}^{+0.036}$ & $2.024$ & $2.198$ \\ 
$10^{+9}K_{5 }$ &$1.933$ & $1.923_{-0.078}^{+0.073}$ & $1.773$ & $2.074$ \\ 
\hline
\multicolumn{5}{|c|}{With CFHTLenS $\qquad -\ln{\cal L}_\mathrm{min} =6776.95$} \\
Param & best-fit & mean$\pm\sigma$ & 95\% lower & 95\% upper \\
\hline 
$100~\omega_{b }$ &$2.235$ & $2.234_{-0.016}^{+0.015}$ & $2.203$ & $2.266$ \\ 
$\omega_{\rm{cdm}}$ &$0.1179$ & $0.1182_{-0.0011}^{+0.0012}$ & $0.116$ & $0.1205$ \\ 
$h$ &$0.686$ & $0.6853_{-0.0054}^{+0.0052}$ & $0.6749$ & $0.6958$ \\ 
$\tau$ &$0.06203$ & $0.05911_{-0.014}^{+0.0085}$ & $0.04$ & $0.07976$ \\ 
$10^{+9}K_{1 }$ &$2.241$ & $2.42_{-0.69}^{+0.67}$ & $1.09$ & $3.784$ \\ 
$10^{+9}K_{2 }$ &$2.294$ & $2.373_{-0.31}^{+0.28}$ & $1.787$ & $2.962$ \\ 
$10^{+9}K_{3 }$ &$2.28$ & $2.292_{-0.063}^{+0.056}$ & $2.176$ & $2.414$ \\ 
$10^{+9}K_{4 }$ &$2.106$ & $2.1_{-0.055}^{+0.035}$ & $2.019$ & $2.189$ \\ 
$10^{+9}K_{5 }$ &$1.899$ & $1.919_{-0.08}^{+0.071}$ & $1.772$ & $2.071$ \\
\hline 
 \end{tabular} \\ 
\caption{Best fit, mean and confidence intervals for the MCMC parameters in the reconstruction with $\alpha_p=0.01$}
\label{tab:FinalResults001}
\end{table}
The degeneracies among the parameters for the PPS value at the knots can be appreciated in the triangle plots of figure~\ref{fig:SWCP_a1_triangle_knots} for $\alpha_p=1$ and figure~\ref{fig:SWCP_a001_triangle_knots} for $\alpha_p=0.01$. Correlations with and among the cosmological parameters not shown are negligible.
As expected, higher penalty induce correlations among the knots which are stronger between neighbouring ones.

Interestingly the only cosmological parameter that correlates with the knots is $\tau_{\rm{reio}}$, which show degeneracy with  the knots at higher $k$ (figure \ref{fig:SWCP_a1_triangle_tau} and \ref{fig:SWCP_a001_triangle_tau}).
This behaviour is however not unexpected.
The $\tau_{\rm{reio}}$ parameter only affects the CMB and in particular its main effect is to suppress the temperature power spectrum at multipoles $\ell \gtrsim 80$.  With our  choice for the location of the knots, the most affected knots are therefore $K_4$ and $K_5$.
Improved polarisation data at low $\ell$ should reduce this degeneracy.
The figure excluding the CFHTLenS dataset is qualitatively very similar and thus is not shown here.

\section{Discussion and conclusions}
\label{sec:conclusions}
The analysis of the  latest cosmological data \cite{Planck:CosmologicalParameters2015} indicates a highly significant  deviation from scale invariance of the primordial power spectrum (PPS)  when parameterized by a power law or by a spectral index and a ``running''.
This offers a powerful tool to  discriminate among theories for the origin of perturbations and among inflationary models. In fact, the deviation from scale invariance of the PPS is a critical prediction of inflation and is the only signature that is generic to all inflationary models. It is therefore a vital test of the inflationary paradigm. 

One may wonder if a strong theory prior on the form of the power spectrum, such as  the  power law prescription, can lead to artificially tight constraints  or even a spurious detection of a deviation from scale invariance, if the adopted model were not a good fit to the data. 

Here we have built on the work of \cite{Sealfon:reconstruction, Verde:MinimallyParamRec2008} to reconstruct the PPS with a minimally parametric approach, using the cross-validation technique as the smoothness criterion. We consider a comprehensive set of state-of-the art cosmological data including probes of the Cosmic Microwave Background,  and of large scale structure via gravitational lensing and galaxy redshift surveys. 
While the spline reconstruction used here is best suited for smooth features in the PPS, it is also sensitive to sharp features if they have high enough signal-to-noise.
 
We find that there is no evidence for deviations from a power law PPS, and that  errors of the reconstructed PPS are comparable with errors obtained with a power law fit.
These results should be compared with those presented in \cite{Peiris:bPlanck2010}, to appreciate the increase in statistical power brought about by the latest generation of experiments.  In fact with current data a scale-invariant power spectrum is highly disfavoured even with this minimally parametric reconstruction. In particular for our conservative choice of smoothness penalty parameter values the  significance of the departure from scale invariance is comparable with that obtained when adopting the ``inflation--motivated'' power-law prior. Constraints no longer relax significantly when  generic forms of the PPS are allowed. 

Because of its flexibility, our reconstruction would be able to detect the tell-tale signature of small scale power suppression induced by free streaming of neutrino if they are sufficiently massive. Of course in reality the suppression happens in the late-time power spectrum, not in the primordial one. But  as we do not include the effect of neutrino masses in the matter transfer function, the reconstruction would recover an ``effective'' small scale damping.  Our reconstruction detects no such signature, ruling out a model with  a power law PPS and sum of neutrino masses of $0.2$ eV or larger.

Our results, which recover in a model independent way  a power law power spectrum with a small but highly significant red tilt,  offer a powerful confirmation of the inflationary paradigm, justifying adoption  of the  inflationary prior in cosmological analyses.

\acknowledgments
LV and AJC acknowledge support by the European Research Council under the European Community's Seventh Framework Programme FP7-IDEAS-Phys.LSS 240117, by  Mineco grant AYA2014-58747-P   and acknowledges Spanish  MINECO MDM-2014-0369 of ICCUB (Unidad de Excelencia 'Mar{\'\i}a de Maeztu').

This work is based on observations obtained with MegaPrime/MegaCam, a joint project of CFHT and CEA/IRFU, at the Canada-France-Hawaii Telescope (CFHT) which is operated by the National Research Council (NRC) of Canada, the Institut National des Sciences de l'Univers of the Centre National de la Recherche Scientifique (CNRS) of France, and the University of Hawaii. This research used the facilities of the Canadian Astronomy Data Centre operated by the National Research Council of Canada with the support of the Canadian Space Agency. CFHTLenS data processing was made possible thanks to significant computing support from the NSERC Research Tools and Instruments grant program.

Funding for the SDSS and SDSS-II has been provided by the Alfred P. Sloan Foundation, the Participating Institutions, the National Science Foundation, the U.S. Department of Energy, the National Aeronautics and Space Administration, the Japanese Monbukagakusho, the Max Planck Society, and the Higher Education Funding Council for England. The SDSS Web Site is http://www.sdss.org/.

The SDSS is managed by the Astrophysical Research Consortium for the Participating Institutions. The Participating Institutions are the American Museum of Natural History, Astrophysical Institute Potsdam, University of Basel, University of Cambridge, Case Western Reserve University, University of Chicago, Drexel University, Fermilab, the Institute for Advanced Study, the Japan Participation Group, Johns Hopkins University, the Joint Institute for Nuclear Astrophysics, the Kavli Institute for Particle Astrophysics and Cosmology, the Korean Scientist Group, the Chinese Academy of Sciences (LAMOST), Los Alamos National Laboratory, the Max-Planck-Institute for Astronomy (MPIA), the Max-Planck-Institute for Astrophysics (MPA), New Mexico State University, Ohio State University, University of Pittsburgh, University of Portsmouth, Princeton University, the United States Naval Observatory, and the University of Washington.

Based on observations obtained with \textit{Planck} (http://www.esa.int/Planck), an ESA science mission with instruments and contributions directly funded by ESA Member States, NASA, and Canada.

\section*{Appendix: Reconstruction sensitivity to non-primordial effects}

\begin{figure}
\centering
\hspace*{\fill}
\begin{subfigure}{.45\textwidth}
\renewcommand\thesubfigure{A.1}
  \centering
  \includegraphics[width=\textwidth]{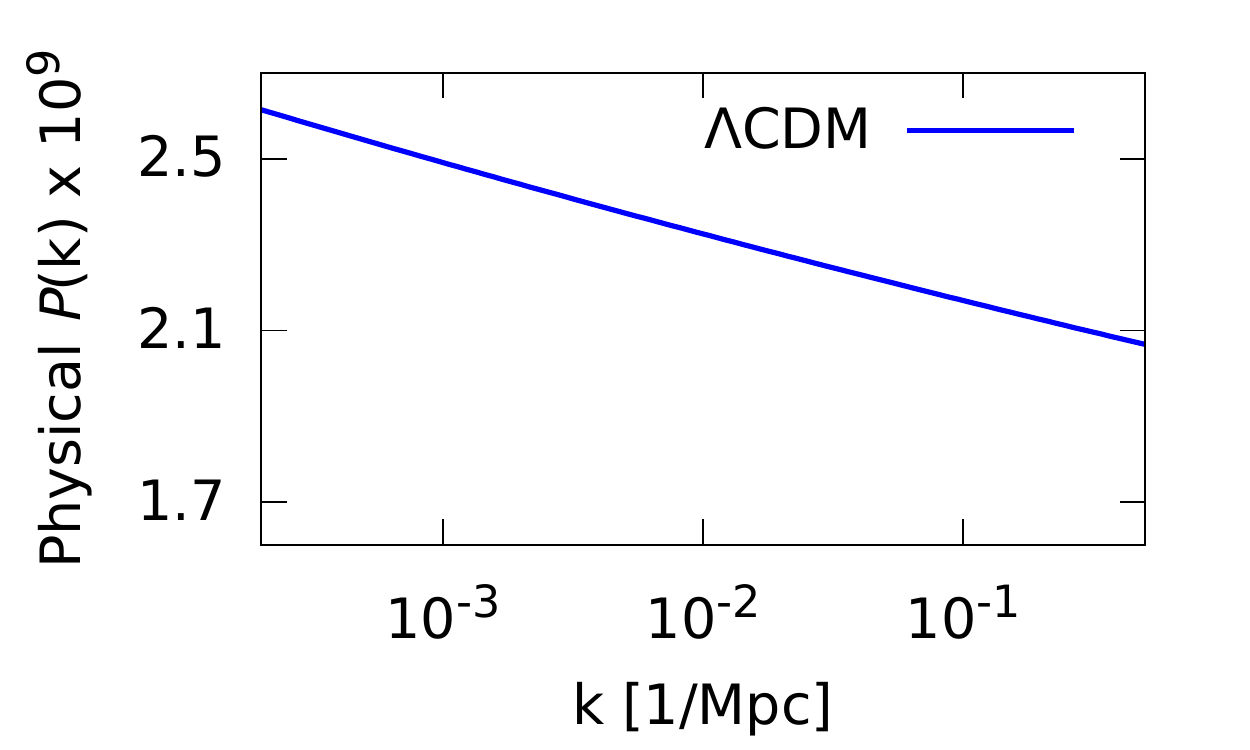}
  \caption{In the first case the initial condition is a power-law PPS.}
  \label{fig:LinearityTrickA1}
\end{subfigure}\hfill%
\begin{subfigure}{.45\textwidth}
\renewcommand\thesubfigure{B.1}
  \centering
  \includegraphics[width=\textwidth]{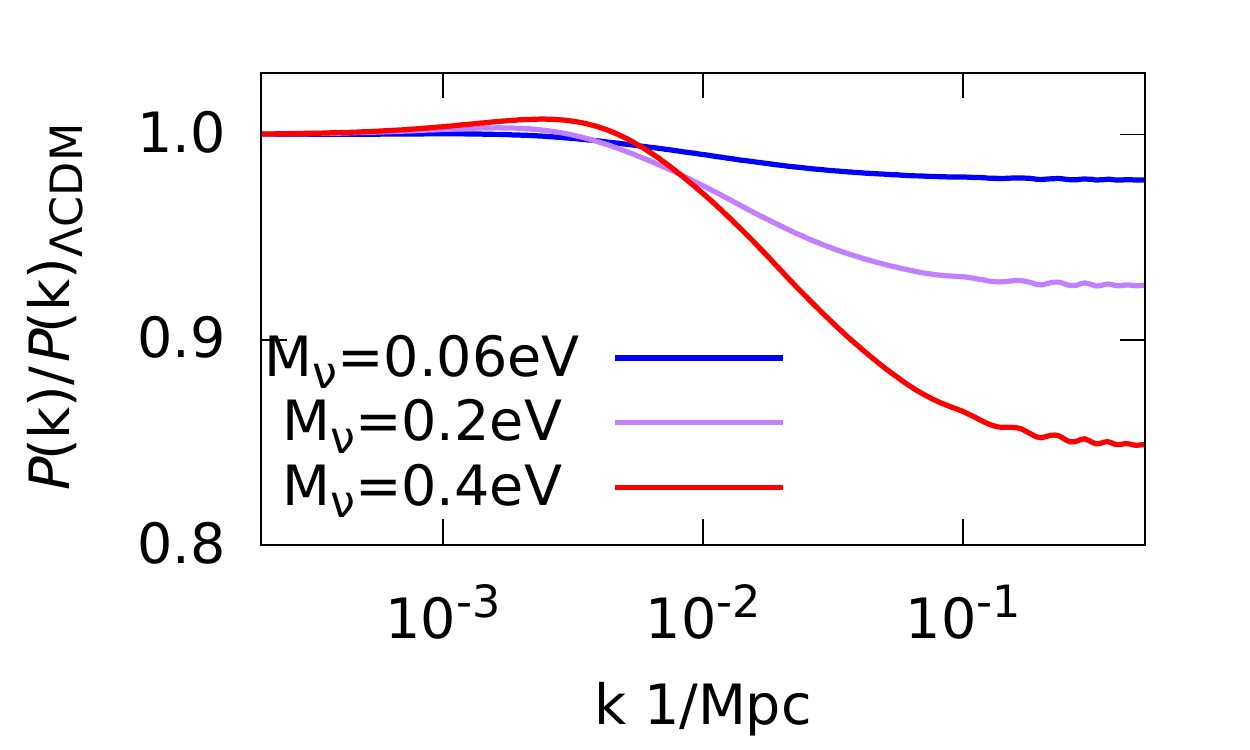}
  \caption{In the second case the neutrino damping is considered in the initial conditions.}
  \label{fig:LinearityTrickB1}
\end{subfigure}
\hspace*{\fill}
\newline
\hspace*{\fill}
\begin{subfigure}{.45\textwidth}
\renewcommand\thesubfigure{A.2}
  \centering
  \includegraphics[width=\textwidth]{"Neutrino_damping_effect"}
  \caption{Then the evolution takes into account the presence of massive neutrino.}
  \label{fig:LinearityTrickA2}
\end{subfigure}
\hfill
\begin{subfigure}{.45\textwidth}
\renewcommand\thesubfigure{B.2}
  \centering
  \includegraphics[width=\textwidth]{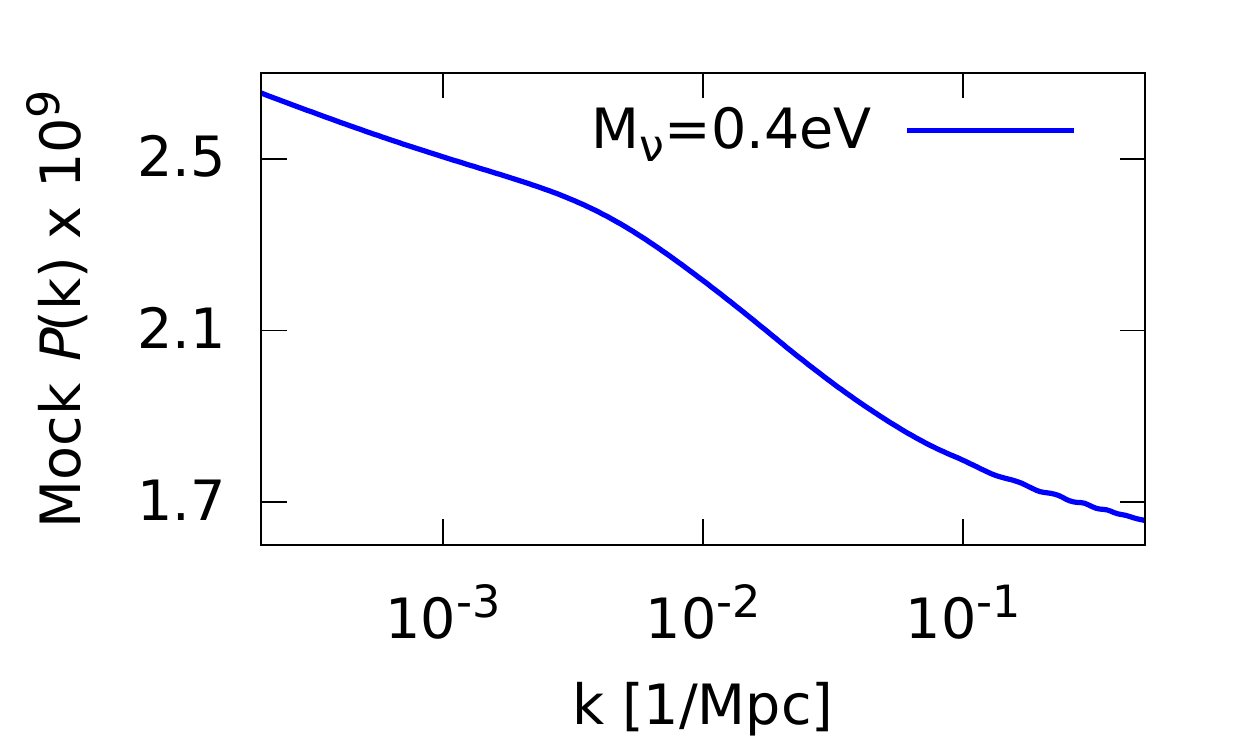}
  \caption{The initial condition is a power-law multiplied by  the neutrino power suppression. Then the evolution equation with massless neutrino is used.}
  \label{fig:LinearityTrickB2}
\end{subfigure}
\hspace*{\fill}
\newline
\hspace*{\fill}
\begin{subfigure}{.45\textwidth}
\renewcommand\thesubfigure{R.1}
  \centering
  \includegraphics[width=\textwidth]{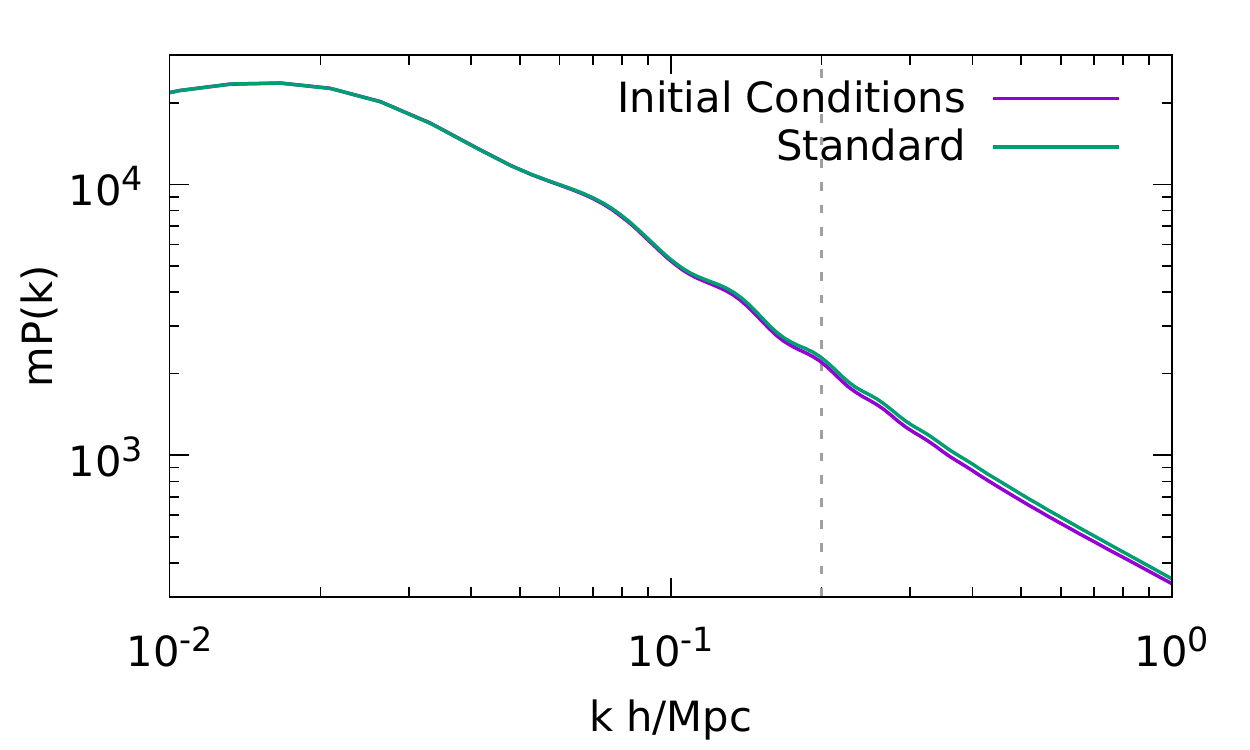}
  \caption{{The two methods generate the same observables in the scales of interest. Here we show the non-linear matter power spectrum.}
  \label{fig:LinearityTrickA3}}
\end{subfigure}
\hfill
\begin{subfigure}{.45\textwidth}
\renewcommand\thesubfigure{R.2}
  \centering
  \includegraphics[width=\textwidth]{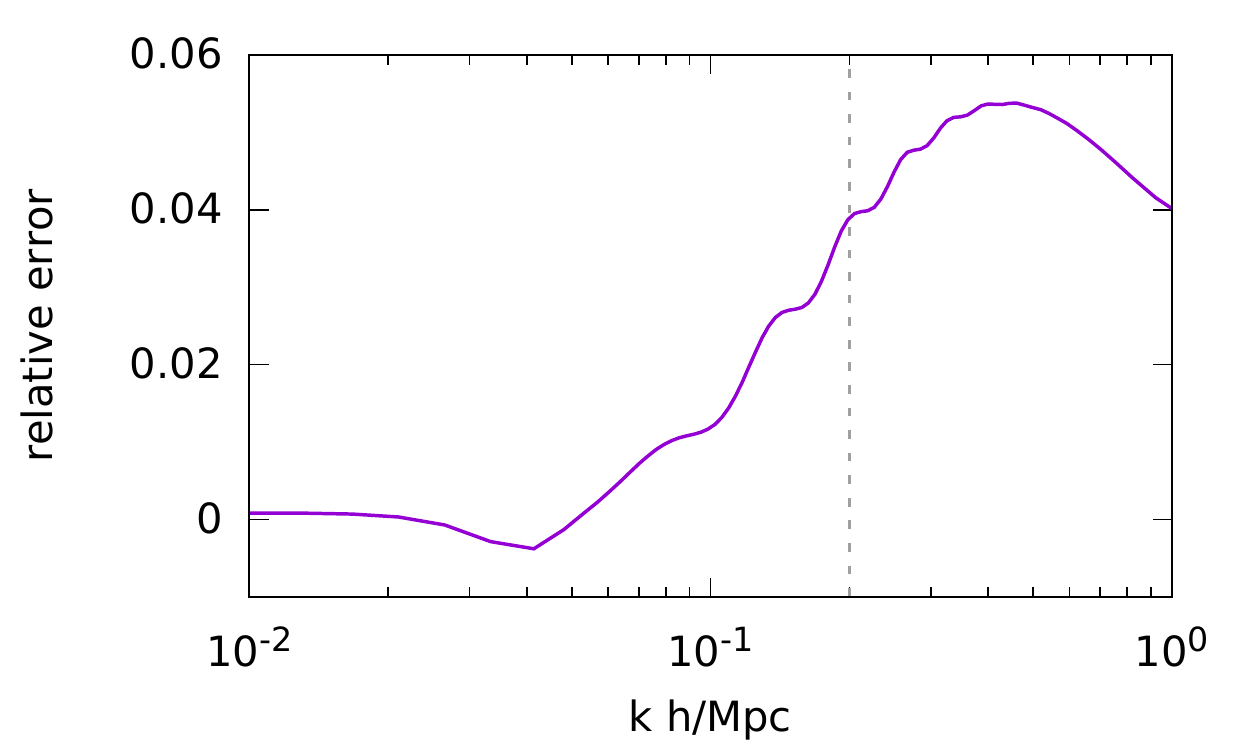}
  \caption{The error do not exceed 1\% at scales larger than  $k=0.1$ $h/$Mpc and is due to non-linearities.}
  \label{fig:LinearityTrickB3}
\end{subfigure}
\hspace*{\fill}
\caption{How a non-zero neutrino mass would induce a power suppression in the reconstructed power spectrum. Both method \subref{fig:LinearityTrickA1}--\subref{fig:LinearityTrickA2} and \subref{fig:LinearityTrickB1}--\subref{fig:LinearityTrickB2} give the same result \subref{fig:LinearityTrickA3}--\subref{fig:LinearityTrickB3} at linear level.}
\end{figure}
Figure \ref{fig:LinearityTrickA1}--\ref{fig:LinearityTrickB3} visualises the  concept --- exploited here --- that in the reconstructed power spectrum the effect of a non-zero neutrino mass is degenerate with a power suppression. This  is a  good approximation especially on scales where the evolution is linear or mildly non-linear, i.e., $k~<~0.2~\ ~h$/Mpc.
Consider a $\Lambda$CDM universe with massive neutrinos, where all the cosmological parameters are known  and with a power law PPS at the end of inflation (figure \ref{fig:LinearityTrickA1}). From these initial conditions we evolve  the perturbations assuming massive neutrino (different values  for the total mass are shown). On small physical scales neutrino free streaming \cite{Lesgourgues:MassiveNuCosmology} suppresses power \ref{fig:LinearityTrickA2}) yielding a resulting power spectrum  shown in figure \ref{fig:LinearityTrickA3}. Now we can think of an alternative method:  implement the neutrino power suppression (figure \ref{fig:LinearityTrickB1}) directly on the initial PPS as a deviation from a power law as shown in figure \ref{fig:LinearityTrickB2}.  This initial power spectrum is then evolved assuming massless neutrinos. The linearity of the perturbation evolution equations guarantees that the generated matter and CMB power spectra would be the same as in the first case (figure \ref{fig:LinearityTrickA3}).
In figure \ref{fig:LinearityTrickB3} we can appreciate the fact that discrepancies in the prediction made in the two cases come from non-linearities.
For example, when considering neutrino with $\Sigma m_{\nu}=0.4$ eV we expect a small scale suppression in the linear power spectrum  of 15\% (figure \ref{fig:LinearityTrickA2}).
The differences due to non linearities  exceed 1\% only above $k=0.1$ $h/$Mpc, and are never more than 4\% at the scales of interest.
This means that in principle this approach we should be able to distinguish the effect for  $\Sigma m_{\nu}\geq 0.06$ eV, even though it would prevent us to  obtain an unbiased  measure in case of detection.

Another source of error that might contribute is given by the use of spline with a limited number of knots.
If the number of knots, or their position, is not suitably chosen, one could be unable to reconstruct a given signal.
It is not our case, with our choice of knots we have verified that we can reconstruct any neutrino power suppression with a $10^{-3}$ accuracy.

\bibliographystyle{JHEP}

\bibliography{bibliografia}
\end{document}